\documentclass[preprint]{elsarticle} 

\makeatletter
	\def\ps@pprintTitle{%
 	\let\@oddhead\@empty
	\let\@evenhead\@empty
	\def\@oddfoot{\centerline{\thepage}}%
	\let\@evenfoot\@oddfoot}
\makeatother

\usepackage{etoolbox}
\patchcmd{\MaketitleBox}{\footnotesize\itshape\elsaddress\par\vskip36pt}{\footnotesize\itshape\elsaddress\par\parbox[b][36pt]{\linewidth}{\vfill\hfill\textnormal{\today}\hfill\null\vfill}}{}{}%
\patchcmd{\pprintMaketitle}{\footnotesize\itshape\elsaddress\par\vskip36pt}{\footnotesize\itshape\elsaddress\par\parbox[b][36pt]{\linewidth}{\vfill\hfill\textnormal{\today}\hfill\null\vfill}}{}{}%

\usepackage[
            colorlinks=true,%
            breaklinks=true,%
            linkcolor=blue,
            urlcolor=blue,%
            citecolor=blue,%
            pdftitle={Title of paper}, 
            pdfkeywords={Keywords},	
            pdfauthor={Authors},		
            bookmarksopen=false,
            pdfpagemode=UseNone]{hyperref}

\usepackage[margin = 1.2in]{geometry}
\usepackage[T1]{fontenc}
\usepackage[english]{babel}
\usepackage[utf8]{inputenc}
\usepackage{hyperref}
\usepackage{tabularx}

\usepackage{mathtools}
\usepackage{amsmath}
\usepackage{amsfonts}
\usepackage{amsthm}
\usepackage{amssymb}
\usepackage{array}
\usepackage{algorithm} 
\usepackage{algorithmicx}
\usepackage{algpseudocode}
\usepackage{subcaption}
\usepackage{siunitx}
\usepackage{arydshln}
\usepackage{nomencl}
\usepackage{framed}  
\usepackage{multicol}  

\usepackage{color}
\usepackage{xcolor}
\usepackage{url}
\usepackage{cleveref}
\usepackage{graphicx}
\usepackage{subcaption}
\usepackage{breakcites}
\usepackage{soul} 
\usepackage{enumitem}
\usepackage[title,titletoc,toc]{appendix}
\usepackage{tikz}
\usetikzlibrary{arrows.meta, positioning}
\usetikzlibrary{patterns}
\usepackage{pgfplots}
\pgfplotsset{compat=1.16}

\usepackage[textsize=tiny]{todonotes}

\definecolor{C00}{rgb}{0.121, 0.467, 0.705} 
\definecolor{C01}{rgb}{1.000, 0.498, 0.055} 
\definecolor{C02}{rgb}{0.172, 0.627, 0.173} 
\definecolor{C03}{rgb}{0.839, 0.153, 0.157} 
\definecolor{C04}{rgb}{0.580, 0.404, 0.741} 
\definecolor{C05}{rgb}{0.549, 0.337, 0.294} 
\definecolor{C06}{rgb}{0.890, 0.467, 0.761} 
\definecolor{C07}{rgb}{0.498, 0.498, 0.498} 
\definecolor{C08}{rgb}{0.737, 0.741, 0.133} 
\definecolor{C09}{rgb}{0.090, 0.745, 0.812} 
\definecolor{darkgreen}{rgb}{0.0, 0.5, 0.0}

{\noindent {\textbf{Proof}:} }%
{\hfill $\Box$ \\[1ex] }

\newcommand{\bit}{\begin{itemize}}
\newcommand{\eit}{\end{itemize}}
\newcommand{\ben}{\begin{enumerate}}
\newcommand{\een}{\end{enumerate}}

\newcommand {\real} {\mathbb{R}}

\newcommand{\bA}{\ensuremath{\mathbf{A}}}

\newcommand{\bD}{\ensuremath{\mathbf{D}}}

\newcommand{\bF}{\ensuremath{\mathbf{F}}}

\newcommand{\bH}{\ensuremath{\mathbf{H}}}
\newcommand{\bI}{\ensuremath{\mathbf{I}}}

\newcommand{\bO}{\ensuremath{\mathbf{O}}}

\newcommand{\bQ}{\ensuremath{\mathbf{Q}}}

\newcommand{\bS}{\ensuremath{\mathbf{S}}}
\newcommand{\bT}{\ensuremath{\mathbf{T}}}

\newcommand{\bV}{\ensuremath{\mathbf{V}}}
\newcommand{\bW}{\ensuremath{\mathbf{W}}}

\newcommand{\bY}{\ensuremath{\mathbf{Y}}}

\newcommand{\ba}{\ensuremath{\mathbf{a}}}

\newcommand{\bc}{\ensuremath{\mathbf{c}}}

\newcommand{\bg}{\ensuremath{\mathbf{g}}}

\newcommand{\bj}{\ensuremath{\mathbf{j}}}

\newcommand{\bp}{\ensuremath{\mathbf{p}}}
\newcommand{\bq}{\ensuremath{\mathbf{q}}}

\newcommand{\bs}{\ensuremath{\mathbf{s}}}

\newcommand{\bv}{\ensuremath{\mathbf{v}}}

\newcommand{\bx}{\ensuremath{\mathbf{x}}}

\newcommand{\cE}{\ensuremath{\mathcal{E}}}

\graphicspath{{./figures/}}

\begin{document}
	
    \begin{frontmatter}
        
        \title{Physically consistent predictive reduced-order modeling by enhancing Operator Inference with state constraints
        }
        
        \author{Hyeonghun Kim\corref{cor1}}
        \ead{hyk049@ucsd.edu}
        \author{Boris Kramer}
        
        \cortext[cor1]{Corresponding author}
        
        \address{Department of Mechanical and Aerospace Engineering, University of California San Diego, CA, United States}
        
        \begin{abstract}
        Numerical simulations of complex multiphysics systems, such as char combustion considered herein, yield numerous state variables that inherently exhibit physical constraints. This paper presents a new approach to augment Operator Inference---a methodology within scientific machine learning that enables learning from data a low-dimensional representation of a high-dimensional system governed by nonlinear partial differential equations---by embedding such state constraints in the reduced-order model predictions. In the model learning process, we propose a new way to choose regularization hyperparameters based on a key performance indicator.
        Since embedding state constraints improves the stability of the Operator Inference reduced-order model, we compare the proposed state constraints-embedded Operator Inference with the standard Operator Inference and other stability-enhancing approaches. For an application to char combustion, we demonstrate that the proposed approach yields state predictions superior to the other methods regarding stability and accuracy. It extrapolates over 200\% past the training regime while being computationally efficient and physically consistent. 

        \end{abstract}	
        
        \begin{keyword}
            Model reduction \sep data-driven modeling \sep nonlinear dynamical system \sep multiscale \sep multiphysics problem \sep Operator Inference \sep state constraints
        \end{keyword}

    \end{frontmatter}

\section{Introduction} \label{sec:intro}
Computational fluid dynamics (CFD) enables simulation of complex multiscale and multiphysics gas-solid flow dynamics in fluidized bed reactors that involve heat and mass transfer, particle collision, as well as homogeneous and heterogeneous chemical reactions~\cite{KU2015270}. However, CFD simulations are computationally expensive, particularly for problems governed by highly nonlinear partial differential equations (PDEs) where fine spatial and temporal discretization is required, which leads to high-dimensional nonlinear systems of ordinary differential equations (ODEs). This poses a significant obstacle for many-query tasks such as optimization, design, and uncertainty quantification. \par

Model reduction provides one avenue to address this challenge by constructing low-dimensional surrogate models for high-dimensional systems. In particular, projection-based model reduction techniques construct a reduced-order model (ROM) by projecting the high-dimensional system onto a low-dimensional subspace, or a nonlinear manifold, and solving the governing equations in the reduced space. This approach effectively incorporates the physics of the high-dimensional problem into the surrogate model; see the textbooks and survey~\cite{quarteroni2014reduced, benner2015survey, hesthaven2016certified}. Such an approach is intrusive because it requires access to codes that implement the semi-discretization of the PDEs of the high-fidelity system, often called the full-order model (FOM). However, in many real-world settings, the solution of highly complex problems often relies on complicated legacy code or commercial software, in which cases access to the FOM operators is not possible, rendering intrusive model reduction not applicable~\cite{ghattas2021learning}. \par

Nonintrusive ROM methods have emerged as an efficient solution to many engineering challenges, leveraging the abundance of data and obviating a need for explicit numerical operators of the FOM.
This paper focuses on Operator Inference (OpInf)~\cite{PEHERSTORFER2016196}, a nonintrusive ROM technique that has been extended in many directions and applied to a wide range of problems in science and engineering over the last decade. OpInf learns a low-dimensional representation of the high-dimensional system by exploiting the physical/mathematical structure and insights from high-fidelity numerical simulation data; see also the recent survey~\cite{kramer2024learning}. OpInf fits the ROM operators to the data by allowing for polynomial terms on the right-hand side of the ROM, effectively capturing the nonlinear dynamics of a system.
To date, OpInf has demonstrated effectiveness in a variety of applications: advection-dominated magnetohydrodynamic systems~\cite{issan2023predicting}, nonlinear aerodynamics and coupled aeroelastic flutter \cite{Zastrow2023}, incompressible flow~\cite{0x003d183f}, Rayleigh and B\'enard convection~\cite{Rocha2023}, and soft robotics~\cite{AdiShaTorKraTol_Soft_Robot_DDROM_SCR2023}, among others. In addition, OpInf for parametric PDEs has been investigated for the shallow water equation~\cite{yildiz2021learning}, the heat equation and the FitzHugh-Nagumo neuron model~\cite{mcquarrie2023nonintrusive}, and a large-scale rotating detonation rocket engine combustion application~\cite{farcas2023parametric}.
However, stability and physical consistency of OpInf ROMs are still open challenges in certain applications, as we discuss below. \par

Numerous studies have been conducted on the stability of ROMs, a necessary condition for maintaining physical consistency. In the intrusive setting, the authors of~\cite{kalashnikova2014stabilization} proposed an approach to
stabilize projection-based ROMs through full-state feedback to modify unstable eigenvalues of a linear operator and through a constrained nonlinear least-squares optimization problem.
In the nonintrusive setting, the authors in~\cite{kaptanoglu2021promoting} adapted the trapping theorem~\cite{schlegel2015long} to promote global stability in data-driven machine learning models with quadratic nonlinearities.
Another avenue to drive systems to stability is adding closure models; see the survey~\cite{ahmed2021closures}. The authors in~\cite{wang2012proper} enhance ROM stability and accuracy by proposing new closure models for POD ROMs of turbulent flows. Furthermore, the authors in~\cite{BBSK17stabilizationROMextremumSeeking, BBK17PODstabilizationBoussinesq} provided a stabilizing closure model for the 2D and 3D Boussinesq equations based on data-driven multi-parametric extremum seeking. Finally, the authors in~\cite{kalb2007intrinsic} introduced a stabilization scheme that leverages the information encapsulated within the POD modes to predict the long-term dynamics of the full-order system accurately. \par

In the OpInf framework, various approaches have been proposed to develop physically consistent ROMs by promoting stability and preserving the physical/mathematical structure present in the FOM. To ensure the stability of OpInf ROMs,~\cite{kramer2021stability} introduced analytical and optimization-based estimates of stability domains for quadratic-bilinear OpInf ROMs using Lyapunov functions. The authors in~\cite{SAWANT2023115836} extended this work to OpInf models with nonlinear cubic terms, where the stability conditions are incorporated as constraints in the OpInf learning problem.
The authors in~\cite{Swischuk_2020} used regularization to obtain a stable OpInf ROM for rocket combustion, and the authors in~\cite{McQuarrie_2021} provided the generalized scalable algorithmic implementation of regularization to improve the robustness of OpInf.
Maintaining physical consistency of ROMs can also be achieved through structure-preserving methods. In particular, embedding the underlying physical and mechanical structure into the OpInf framework has been extensively studied via structure-preserving OpInf for canonical Hamiltonian systems~\cite{SHARMA2022133122}, for noncanonical Hamiltonian systems~\cite{GRUBER2023116334}, for Lagrangian systems~\cite{SHARMA2024134128, SHARMA2024116865}, for general systems that are conservative or dissipative in energy, and which exhibit a gradient structure~\cite{GENG2024117033}, and for linear mechanical systems with second-order structure~\cite{filanova2023operator}. Lastly, the authors in~\cite{koike2024energy} enforced the energy-preserving structure in the quadratic operator of OpInf through constrained optimization. \par

Physical systems often exhibit inherent bounds on their state variables, imposing crucial constraints that must be respected for accuracy in the FOM or ROM modeling. For example, species concentrations in chemical systems must remain non-negative and often have maximum possible concentrations due to constraints posed by reaction equations and boundary conditions. Fluid velocities in specific flow regimes cannot exceed a particular value (e.g., the speed of sound in the subsonic regime), and temperatures in thermodynamic systems may have lower or upper bounds depending on boundary conditions, etc. Violating these constraints, posed by fundamental physical laws of the system of interest, can lead to unphysical behavior of states in ROMs, poor prediction accuracy, and often model instability. 
Although the aforementioned OpInf ROM approaches conserve mathematical stability (e.g., Lyapunov), system energy, and mechanistic principles, they do not necessarily guarantee that the ROM predictions are consistent with the inherent physical bounds.
In intrusive projection-based ROMs of reacting flows, the authors of~\cite{huang2020species} embedded state constraints on species concentrations to address unphysical oscillations observed in certain quantities and spatial regions and obtained stable ROMs within the training time interval. These oscillations were attributed to the ROM's sensitivity to the coupling between fluctuations in species mass fractions and temperature. To mitigate this issue, they used conditional probability density functions, which improved the prediction robustness. However, when reliable conditional PDFs between certain states are not available, this approach does not apply.
Moreover, while previous OpInf formulations in~\cite{McQuarrie_2021, Swischuk_2020, qian2022reduced, jain2021performance} have successfully avoided stability issues related to unphysical predictions in a single-injector rocket combustion application, a general framework to systematically preserve inherent state bounds within the state prediction process is still lacking. \par

In this paper, we propose a new method to incorporate state constraints into a nonintrusive ROM, enabling it to produce physically consistent predictions for complex nonlinear systems. 
Our approach builds on the concept of species limiters introduced in~\cite{huang2020species} for intrusive projection-based ROMs of reacting flows, but generalizes it to the nonintrusive and predictive setting. The proposed methodology departs from~\cite{huang2020species} in three key aspects:~(i) state constraints are explicitly enforced during the ROM prediction step via an online correction mechanism that partially reintroduces FOM information, thereby establishing a hybrid offline-online framework, (ii) the framework applies beyond the training regime, and (iii) it is general, i.e., it can be applied to any ROM that evolves low-dimensional states with known physical bounds of high-dimensional states—not solely to species concentrations.
Beyond these three advances, we further propose a new way of selecting regularization hyperparameters in ROMs via application-specific key performance indicators (KPIs). While illustrated in this work using thermal energy in char combustion, this approach can be applied to other reacting flow problems or physical systems, allowing regularization to target the most relevant system metrics for accuracy and stability.
\par
The rest of the paper is organized as follows.~\Cref{sec:char_combustion_application} presents the theoretical and simulation background of char combustion.~\Cref{sec:OpInf_with_SL} introduces the OpInf ROM with state constraints and also compares stabilizing effects provided by two different types of bases.~\Cref{sec:numerical_results} provides numerical results, comparing the proposed method with the standard OpInf and other stability-enhancing approaches, and~\Cref{sec:conclusion} gives conclusions and an outlook to future work.

\section{Char combustion application} \label{sec:char_combustion_application}
A fluidized bed combustor is a combustion technology characterized by its efficient burning of fuels, such as char and biomass, with low pollutant emissions. It combines intricate processes characterized by a multiscale flow structure, the multiphysics associated with dense two-phase reactive flows, a nonlinear coupling of gas and solids, as well as complex homogeneous and heterogeneous chemical reactions involved~\cite{geng2011extended, yang2019particle}. We simulate char combustion with a numerical model via the particle-in-cell (PIC) method, using the open-source software MFiX-PIC (version 23.1.1)~\cite{osti_1630426}.~\Cref{ss:governingEq} presents some theoretical background for the MFiX-PIC, briefly describing the governing equations for the gas and solid phase models alongside the associated chemical equations of the char combustion model.~\Cref{ss:compDomain} defines the computational domain, parameters, and initial and boundary conditions of the char combustion problem.~\Cref{ss:discretization_detail} presents details of the state variables in the MFiX-PIC data, and~\Cref{ss:convergence_study} presents a convergence study of the char combustion simulation.

\subsection{Governing equations of MFiX-PIC} \label{ss:governingEq}
In the PIC method, the term \textit{particle} refers to an individual unit of material characterized by specific physical properties such as density, diameter, and chemical composition. The PIC method groups particles with the same physical properties as a computational parcel, a statistical assembly of particles. The PIC method solves a two-phase fluid problem by considering a Lagrangian solids phase model that tracks the trajectory of each computational parcel, not each particle, within an Eulerian fluid. Thus, it achieves considerable efficiency in computational costs compared to other methods that track each particle. Here, we briefly present the governing equations of the Eulerian gas phase model and the Lagrangian solids phase model that are solved in MFiX-PIC, as in~\cite{osti_1630426, osti_1604993}.

\subsubsection{Eulerian gas phase model}    \label{sss:Eulerian_gas_phase}
The Eulerian gas phase model is governed by conservation equations of the gas phase mass, $i$th species mass in the gas mixture, momentum, and internal energy, constituted mathematically as:
\begin{align}
    \frac{\partial}{\partial t}(\varepsilon_{g}\rho_{g}) + \nabla \cdot (\varepsilon_{g} \rho_{g} \bv_{g}) &= \sum_{i=1}^{N_{g}} R_{g,i} + S_{g} \label{eq:gas_mass_conservation}, \\
    \frac{\partial}{\partial t}(\varepsilon_{g}\rho_{g}Y_{i}) + \nabla \cdot (\varepsilon_{g} \rho_{g} \bv_{g} Y_{i}) &= \nabla \cdot (\varepsilon_{g} \bj_{g}) + R_{g,i} + S_{g,i} \label{eq:gas_species_mass_conservation}, \\
     \frac{\partial}{\partial t}(\varepsilon_{g}\rho_{g} \bv_{g}) + \nabla \cdot (\varepsilon_{g} \rho_{g} \bv_{g} \bv_{g}) &= - \nabla p_{g} + \nabla \cdot \boldsymbol{\tau}_g + \varepsilon_{g} \rho_{g} \bg + \bS_{g} \label{eq:gas_momemtum_conservation}, \\
     \varepsilon_{g}\rho_{g}c_{p,g} \left [ \frac{\partial T_{g}}{\partial t} + \nabla \cdot ( \bv_{g} T_{g}) \right ] &= \nabla \cdot \left( \varepsilon_{g} \kappa \nabla T_{g} \right) + S_{g} \label{eq:gas_energy_conservation}.
\end{align}
In Eq.~\eqref{eq:gas_species_mass_conservation}, the diffusive species mass flux vector $\bj_{g}$ is determined by Fick's law $\bj_{g} = -\rho_{g} D_{g,i} \nabla Y_{i}$, where $D_{g,i}$ is the diffusion coefficient.
The scalar $S_g$ is a user-defined total gas-phase mass source [$\text{kg}/(\text{m}^3\cdot\text{s})$], with $S_{g,i}$ denoting the contribution from $i$th species, so that $S_g = \sum_i S_{g,i}$, assuming all species contributions are accounted for. The vector $\mathbf{S}_g$ is a general gas-phase momentum source [$\text{kg}/(\text{m}^2\cdot\text{s}^2)$], accounting for interphase momentum exchange.
In Eq.~\eqref{eq:gas_momemtum_conservation}, the stress tensor $\boldsymbol{\tau}_g$ is modeled in the gas phase as Newtonian fluids with a linear relation between stress and rate of strain, $\boldsymbol{\tau}_g = - \frac{2}{3} \mu_{g} (\nabla \cdot \bv_{g}) \boldsymbol{\delta} + \mu_{g} \left [ \left ( \nabla \bv_{g} \right) + \left ( \nabla \bv_{g} \right )^{\top} \right]$. Here, $\mu_{g}$ is the gas dynamic viscosity [$\textnormal{kg/(m} \cdot \textnormal{s)}$], and $\boldsymbol{\delta}$ represents the Kronecker delta tensor, characterized by elements $\delta_{j,k} = 0$, if $j \neq k$, and $\delta_{j,k} = 1$, if $j = k$. Additional notations are defined in the nomenclature in~\ref{app:nomenclature}.

\subsubsection{Lagrangian solids phase model}    \label{sss:Lagrangian_solids_phase}
The Lagrangian solids phase model is governed by the conservation of mass, $i$th species mass, momentum, and internal energy for the $p$th parcel as:
\begin{align}
    \frac{\rm d}{\mathrm{d}t}(W_{p} m_{p}) &= W_{p} \sum_{i=1}^{N_{g}} R_{p,i} \label{eq:solid_mass_conservation}, \\
    \frac{\rm d}{\mathrm{d}t}(W_{p} m_{p} Y_{p,i}) &= W_{p} R_{p,i} \label{eq:solid_species_conservation}, \\
    W_{p} m_{p} \frac{\rm d \bv_{p}}{\mathrm{d}t} &= W_{p} \left( m_{p} \bg + \frac{m_{p}}{\varepsilon_{s} \rho_{s}} \nabla \cdot \boldsymbol{\tau}_{p} \right ) \label{eq:solid_momentum_conservation}, \\
    W_{p} m_{p} c_{p,p} \frac{\mathrm{d} T_{p}}{\mathrm{d}t} &= - W_{p} \left( \sum_{i=1}^{N_{g  }} h_{p,i} R_{p,i} \right ) + S_{p}. \label{eq:solid_energy_conservation}
\end{align}
The solid volume fraction is $\varepsilon_s = 1 - \varepsilon_g$, and $\rho_s$ represents the solid density. Note that the statistical weight of particles $W_p$ is considered in Eqs.~\eqref{eq:solid_mass_conservation}--\eqref{eq:solid_energy_conservation}. \par

\subsubsection{Chemical reactions}  \label{sss:chemical_eqs}
In the char combustion, the heterogeneous gas-solid and homogeneous gas-gas chemistry are considered. First, the heterogeneous reaction occurs under the effect of char that reacts with oxygen, producing carbon monoxide as
\begin{align}
    \textnormal{C} + 0.5\textnormal{O}_2 \longrightarrow \textnormal{CO}.
\label{eq:heterogenous reaction}
\end{align}
This \textit{char combustion} process, where char is burned as a fuel, consists of three distinct phases~\cite{winter1997temperatures, basu2006combustion}. Initially, when a large char particle enters the bed, the reaction rate significantly exceeds the bulk diffusion rate. In this phase, bulk diffusion governs the combustion process. As combustion progresses, the char particle shrinks, making the diffusion rate comparable to the reaction rate. Lastly, the char particles become so small that the diffusion rate far exceeds the kinetic rate, making the kinetic reaction rate an influential factor in determining the combustion rate.
The diffusion rate $R_{\textnormal{diff}}$ and the kinetic reaction rate $R_{\textnormal{chem}}$ are computed following~\cite{geng2011extended, XIE202120} as
\begin{align*}
    R_{\textnormal{diff}}=\frac{24 Sh D_{o}}{d_{p} R_u T_{m}}, \hspace{1.5em} R_{\textnormal{chem}} = A_{\textnormal{pre}} \textnormal{exp} \left(- \frac{E_{0}}{R_u T_{p}} \right).
\end{align*}
Considering the interaction between $R_\textnormal{diff}$ and $R_{\textnormal{chem}}$, the char combustion rate [$\textnormal{kmol}/\textnormal{s}$] across the aforementioned three phases is determined as
\begin{align*}
    \frac{\mathrm{d} m_{\textnormal{ci}}}{\mathrm{d}t} = -\pi d_{p,i}^2 p_{\textnormal{oxy}} \left ( \frac{1}{R_{\textnormal{diff}}} + \frac{1}{R_{\textnormal{chem}}} \right ) ^{-1}.
\end{align*}
Here, the char particle diameter $d_{p,i}$ is computed as $\left( \frac{6m_p}{\pi \rho_p} \right)^{1/3}$, and the partial pressure of oxygen in the gas mixture $p_{\textnormal{oxy}}$ is computed as $p_g Y_{\textnormal{O}_2} M_{\textnormal{mix}}/M_{\textnormal{O}_2}$.
To compute the Sherwood number, we take the Fr\"{o}ssling-type correlation for the mass transfer of particles in a fluidized bed as
$ Sh = 2.0 \varepsilon_{g} + 0.7 (Re_{g}/\varepsilon_{g})^{1/2} \cdot Sc^{1/3} $,
where the Reynolds number $Re_{g} = \rho_{g} \overline{U} d_{p}/\mu_{g}$, and the Schmidt number $Sc = \mu_{g}/(\rho_{g} D_o)$; see~\cite{LANAUZE19841623, SCALA20074159}.
The homogeneous reaction is the CO combustion, governed by
\begin{align}
    \textnormal{CO} + 0.5\textnormal{O}_2 \longrightarrow \textnormal{CO}_2.
\label{eq:homogeneous reaction}
\end{align}
The reaction rate of the CO combustion is modeled as in~\cite{DRYER1973987} as
\begin{equation}
    r_{\textnormal{CO}} = 3.98 \times 10^{14} \textnormal{exp} \left( - \frac{1.67 \times 10^5}{R_u T_g} \right) [c_{\textnormal{CO}}] [c_{\textnormal{H}_2\textnormal{O}}]^{0.5} [c_{\textnormal{O}_2}]^{0.25}.
\label{eq:CO_combustion_rate}
\end{equation}
Note, since Eq.~\eqref{eq:CO_combustion_rate} uses $c$ (molar concentration) with a unit of [$\textnormal{mol}/\textnormal{cm}^3$], the resulting rate of CO combustion $r_{\textnormal{CO}}$ must be converted to be a unit consistent with the char combustion rate of [$\textnormal{kmol}/\textnormal{s}$].

\subsection{Computational domain of the char combustion simulation} \label{ss:compDomain}
\begin{figure}[H]
    \centering
    \includegraphics[width=0.3\linewidth]{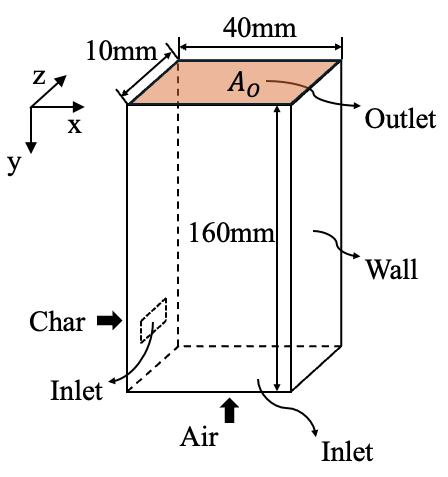}
    \caption{Computational domain of a laboratory-scale fluidized bed for char combustion.
    }
    \label{fig:computational_domain}
\end{figure}

The computational domain of the fluidized bed boiler for char combustion is shown in~\Cref{fig:computational_domain}. The small inlet dotted on the left wall of the boiler ($yz$-plane, with $x=0[\textnormal{mm}]$) is fed with solid char that has a constant mass flow rate of $5.921\times10^{-6} [\textnormal{kg}/\textnormal{m}]$, and the entire bottom inlet ($xz$-plane, with $y=160[\textnormal{mm}]$) is fed with air. The boiler outlet $A_O$ ($xz$-plane, with $y=0[\textnormal{mm}]$) has a constant atmospheric pressure. The physical parameters and initial/boundary conditions used for the simulation are tabulated in~\Cref{tab:combustion_parameters}. We set a slightly lower initial background temperature than the inflowing char temperature to obtain a transient thermal ramp-up during the initial few seconds of operation. Our simulation setup is based on the approaches presented in~\cite{lee2024global} and~\cite{XIE202120}.
    
\begin{table}[H]
    \centering
    \caption{Physical properties and parameters used for the char combustion simulation.}
    \vspace{-0.5em}
    \begin{tabular}{l l}
        \hline
        Parameters & Values \\
        \hline
        Dimensions, $x \times y \times z$ [mm] & $40 \times 160 \times 10$ \\
        Char inflow diameter [\textnormal{m}] & $4.906 \times 10^{-4}$ \\
        Char inflow density [$\textnormal{kg}/\textnormal{m}^3$] & $1{,}300$ \\
        Char mass flow rate [$\textnormal{kg}/\textnormal{s}$] & $5.921  \times 10^{-6}$ \\
        Char inflow temperature [$\textnormal{K}$] & $300$ \\
        Sand initial diameter & $3.705  \times 10^{-4}$ \\
        Sand density [$\textnormal{kg}/\textnormal{m}^3$] & $2{,}650$ \\
        Initial background gas temperature [$\textnormal{K}$] & $1{,}073$ \\
        Initial background solid temperature [$\textnormal{K}$] & $293.150$ \\
        Bottom inlet gas pressure [$\textnormal{Pa}$] & $101{,}325$ \\
        Bottom inlet gas temperature [$\textnormal{K}$] & $1{,}073$ \\
        Bottom inlet gas $y$-velocity [$\textnormal{m}/\textnormal{s}$] & $-0.873$ \\
        Bottom inlet $\textnormal{O}_2$ mass fraction & $0.230$ \\
        Bottom inlet $\textnormal{N}_2$ mass fraction & $0.760$ \\
        Bottom inlet $\textnormal{H}_2\textnormal{O}$ mass fraction & $0.010$ \\
        Outlet pressure [\textnormal{Pa}] & $101{,}325$ \\
        Boundary wall temperature [$\textnormal{K}$] & $1{,}073$ \\
        \hline
    \end{tabular}
    \label{tab:combustion_parameters}
\end{table}

\subsection{MFiX-PIC discretization details and dataset} \label{ss:discretization_detail}
To simulate the char combustion process, MFiX-PIC employs the finite volume method to spatially discretize the governing PDEs~\eqref{eq:gas_mass_conservation}--\eqref{eq:solid_energy_conservation}. This results in an $N$-dimensional system of nonlinear ODEs in terms of the full state $\bq = \bq(t) \in \real^N$, describing the time evolution of the $n_v$ variables with
\begin{equation}
    \frac{\mathrm{d}\bq}{\mathrm{d}t} = \bF(\bq), \hspace{1.5em} \bq(t_0)=\mathbf{q}_0
\label{eq:general_nonlinear_ODE}
\end{equation}
for $0 < t \leq T_f$. Due to the constant boundary conditions in the combustion model, as shown in~\Cref{tab:combustion_parameters}, the nonlinear system represented by Eq.~\eqref{eq:general_nonlinear_ODE} is autonomous. We choose the $n_v = 10$ gas phase primitive variables---$p_g$, $v_x$, $v_y$, $v_z$, $T_g$, and $Y_i$, where $i=1$ for $\textnormal{O}_2$, $i=2$ for $\textnormal{N}_2$, $i=3$ for $\textnormal{CO}$, $i=4$ for $\textnormal{CO}_2$, and $i=5$ for $\textnormal{H}_2\textnormal{O}$---as solution variables.
In the time-discretized version of Eq.~\eqref{eq:general_nonlinear_ODE}, the full state $\bq_k$ approximates $\bq(t_k)$ for each time step $k$, and it includes all the solution vectors as:
\begin{align}
    \bq_k = \left[\bp_k^{\top} \hspace{0.5em} \bv_{x,k}^{\top} \hspace{0.5em} \bv_{y,k}^{\top} \hspace{0.5em} \bv_{z,k}^{\top} \hspace{0.5em} \bT_k^{\top} \hspace{0.5em} \bY_{\textnormal{O}_2, k}^{\top} \hspace{0.5em} \bY_{\textnormal{N}_2, k}^{\top} \hspace{0.5em} \bY_{\textnormal{CO}, k}^{\top} \hspace{0.5em} \bY_{\textnormal{CO}_2, k}^{\top} \hspace{0.5em} \bY_{\textnormal{H}_2\textnormal{O}, k}^{\top} \right]^{\top} \in \real^N.
\label{eq:solution_variables}
\end{align}
Here, $N = n_v \cdot n_x$, and the bold symbol elements in $\bq_k$ indicate the vectorized solutions corresponding to each solution variable on the discretized $n_x$ cells. Denote the $\ell$th state vector within $\bq_k$ as
\begin{align}
    \bq_k^{(\ell)} = \bq_k\left[ (\ell-1)n_x: \ell n_x - 1 \right], \hspace{1em} \bq_k^{(\ell)} \in \real^{n_x}.
\label{eq:l_th_state}
\end{align}
Here, $\ell \in \{1,2,\dots,n_v\}$, and $(\ell-1)n_x: \ell n_x - 1$ indicates the range of indices pertaining to the $\ell$th variable in the full state $\bq_k$. The index $\ell$ in our case follows the order of variables in Eq.~\eqref{eq:solution_variables}, e.g., $\ell=1$ for pressure, $\ell=2$ for $x$ velocity, $\cdots$, and $\ell=10$ for $\textnormal{H}_2\textnormal{O}$ mass fraction.
The training snapshot matrix is constructed as
\begin{equation}
    \bQ = \left[ \bq_0 \hspace{1em} \bq_1 \hspace{0.5em} \dots \hspace{0.5em} \bq_{K-1} \right] \in \real^{N \times K}
\label{eq:training_snapshot}
\end{equation}
where the subscript of $\bq$ represents the time step and $K$ denotes the number of snapshots in the training regime. Later, we denote each $\ell$th state snapshot matrix as $\bQ^{(\ell)} \in \real^{n_x \times K}$. This notation is frequently used in~\Cref{ss:data_pre_processing}.

\subsection{Convergence study of the char combustion simulation} \label{ss:convergence_study}
We conduct a convergence study to verify that the numerical simulation of the char combustion provides reliable solutions that become independent of the spatial discretization parameter $n_x$.
For the quantity of interest of the analysis, we choose the time-integrated species mass fraction that is averaged at the boiler outlet, denoted as $\overline{\bY}$, of $\textnormal{O}_2$ and $\textnormal{CO}_2$, defined as
\begin{align}
    \overline{\bY} = \sum_{k} \left( \frac{1}{|M|}\sum_{j \in M} \bY_{j,k} \right),
\label{eq:integrated_top_mass_fraction}
\end{align}
where $\bY_{j,k}$ denotes the mass fraction in the spatial index $j$ and the temporal index $k$. Here, $M$ is the set of spatial indices corresponding to the boiler outlet $A_O$ (see~\Cref{fig:computational_domain}), and $|M|$ is the cardinality of that set. The quantity of interest $\overline{\bY}$ provides a comprehensive understanding of the mass fractions within the region designated for thermal energy collection, integrated over time. We consider the simulation over a temporal interval ranging from 0 to 2.5 seconds, after which the variables reach a quasi-steady state.
~\Cref{fig:convergence_study_plot} shows the quantity of interest versus the spatial discretization dimension $n_x$, as well as the computation time required, measured in CPU hours, for each $n_x$.
The results for different mesh sizes are saved with the constant time step of $10^{-3}$s, and the same final time is considered. Hence, the difference in values of $\overline{\mathbf{Y}}$ arises only from the differences in the species mass fractions. For increasing mesh sizes $n_x$, the observed values of $\overline{\bY}_{\textnormal{O}_2}$ are 340, 395, 439, 440, and 468, while the corresponding $\overline{\bY}_{\textnormal{CO}_2}$ values are 236, 166, 111, 110, and 80. These results indicate that the integrated mass fractions of $\textnormal{O}_2$ and $\textnormal{CO}_2$ begin to converge at a spatial discretization of $n_x = 22{,}400$, making further mesh-refinement not cost-effective. Consequently, we select $n_x=22{,}400$ as the spatial discretization dimension, leading to an overall ODE dimension of $N = n_v \cdot n_x = 224{,}000$.
Nevertheless, as shown in~\Cref{fig:convergence_study_plot}, the FOM simulation with the chosen mesh size $22{,}400$ still requires an extensive amount of computational cost, over $4{,}000$ CPU hours, to obtain data for a physical time of 2.5 seconds.
\begin{figure}[H]
\centering
\begin{tikzpicture}
\begin{axis}[
    width=0.8\textwidth,
    height=0.3\textwidth,
    xlabel=Mesh size $n_x$,
    ylabel={$\overline{\bY}$},
    xmin=0, xmax=92000,
    xtick={2520,10800,22400,42000,88200},
    xticklabels={2520,10800,22400,42000,88200},
    ytick={100,200,300,400},
    ticklabel style={/pgf/number format/fixed},
    xmajorgrids=true,
    ymajorgrids=true,
    grid style=dashed,
    scaled ticks=false,
    legend style={at={(0.97,0.70)}, anchor=north east},
    ytick style={red},
    yticklabel style={red},
]
\addplot[
    red,
    mark=square,
    thick
] coordinates {
    (2520,340)
    (10800,395)
    (22400,439)
    (42000,440)
    (88200,468)
}; 
\addplot[
    red,
    mark=square,
    dashed,
    thick
] coordinates {
    (2520,236)
    (10800,166)
    (22400,111)
    (42000,110)
    (88200,80)
};
\legend{$\textnormal{O}_2$, $\textnormal{CO}_2$}
\end{axis}

\begin{axis}[
    width=0.8\textwidth,
    height=0.3\textwidth,
    xmin=0, xmax=92000,
    ylabel={CPU hours (log scale)},
    ymin=100, ymax=100000,
    ymode=log,
    axis x line=none,
    axis y line*=right,
    legend style={at={(0.97,0.35)}, anchor=north east},
    ytick style={blue},
    yticklabel style={blue},
]
\addplot[
    blue,
    dash dot,
    mark=*,
    thick
] coordinates {
    (2520,405)
    (10800,1020)
    (22400,4320)
    (42000,7560)
    (88200,25000)
};
\legend{CPU hours}
\end{axis}
\end{tikzpicture}

\caption{Convergence study. Mesh sizes $n_x \in \{2520, 10800, 22400, 42000, 88200\}$ are evaluated. The left y-axis corresponds to integrated mass fractions for $\textnormal{O}_2$ and $\textnormal{CO}_2$. The right y-axis corresponds to CPU hours on a logarithmic scale.}
\label{fig:convergence_study_plot}
\end{figure}

\section{Operator Inference reduced-order model with state constraints} \label{sec:OpInf_with_SL}
The char combustion model is governed by the gas phase conservation equations~\eqref{eq:gas_mass_conservation}--\eqref{eq:gas_energy_conservation}, the solid phase conservation equations~\eqref{eq:solid_mass_conservation}--\eqref{eq:solid_energy_conservation}, and the chemical reactions~\eqref{eq:heterogenous reaction}--\eqref{eq:homogeneous reaction}. However, similar to other legacy codes or commercial software, the complexity in the MFiX-PIC source code renders the derivation of high-fidelity operators associated with the nonlinear function $\bF$ in Eq.~\eqref{eq:general_nonlinear_ODE} impractical.
For exactly those settings, OpInf has emerged as an efficient scientific machine learning technique to construct predictive ROMs of high-dimensional dynamical systems. In particular, OpInf solely requires high-fidelity data to develop underlying low-fidelity models by learning polynomial ROMs. \par
In~\Cref{ss:data_pre_processing} and~\Cref{ss:dimensionality_reduction}, we present data pre-processing and dimension reduction techniques essential for constructing ROMs of multiphysics, multiscale applications.~\Cref{ss:OpInf_regularization} provides the theoretical background of the standard OpInf with regularization. In~\Cref{ss:species_limiter}, we introduce state constraints and their implementation details in the OpInf ROM. We also illustrate the stabilizing effect of state constraints associated with the use of a global POD basis and block-structured POD basis.

\subsection{Data pre-processing} \label{ss:data_pre_processing}
The training and testing datasets are drawn from the FOM simulation data, with $t\in[2,14]$s used for training and $t\in[14,40]$s reserved for testing.
The variables in the training dataset encompass a broad range of numerical scales, as shown in~\Cref{tab:range_of_variable}. Specifically, pressure and temperature are on the orders of $10^5$ and $10^3$, respectively, whereas the mass fractions of certain species have magnitudes around $10^{-4}$. Such multiphysics and multiscale problems pose substantial challenges for model training, as larger magnitude variables can dominate those with smaller scales. To avoid numerical issues, it is advantageous to pre-process the FOM data to scale all variables to be in comparable orders of magnitude.
Among the state variables, pressure and temperature (including their mean values) have relatively large magnitudes among all the variables considered. To address this, we center each pressure and temperature snapshot around zero via mean subtraction: 
\begin{align}
    \bq^{(\ell),{\textnormal{shifted}}}_{k} = \bq^{(\ell)}_{k} - \overline{\bq}^{(\ell)}, \hspace{2em} \overline{\bq}^{(\ell)} = \frac{1}{K} \sum_{k=0}^{K-1} \bq^{(\ell)}_{k},
\label{eq:shifting}
\end{align}
where $\ell$ indicates the state indices as in Eq.~\eqref{eq:l_th_state}, $\bq^{(\ell)}_{k} \in \real^{n_x}$ is the $k$th snapshot of the $\ell$th state, and $\overline{\bq}^{(\ell)} \in \real^{n_x}$ is its mean over $K$ snapshots. We denote the shifted snapshot matrix as $\bQ^{(\ell), \textnormal{shifted}} \in \real^{n_x \times K}$, populated by the centered snapshots $\bq^{(\ell),\textnormal{shifted}}_{k}$, where $k=0,1,\dots,K-1$.
In contrast, the $x$, $y$, and $z$ velocities have relatively smaller magnitudes, with their means close to zero. Thus, we do not shift them. 
After shifting, if necessary, the pressure, temperature, and velocity snapshots are then scaled to ensure that all variables fall within the range of [$-1,1$]. We scale variables by dividing all elements by the maximum absolute value of the entire shifted matrix of each variable $\bQ^{(\ell),\textnormal{shifted}}$. The scaling process results in the scaled snapshot matrix of the $\ell$th variable as:
\begin{align}
    \bS^{(\ell)} = \frac{\bQ^{(\ell),{\textnormal{shifted}}}}{\textnormal{max}\left( | \bQ^{(\ell),\textnormal{shifted}} | \right)} \in \real^{n_x \times K}.
\label{eq:scaling}
\end{align}
We adopt this nondimensionalization approach from~\cite{Swischuk_2020}, where mapping variables to the range [$-1,1$] was chosen after careful study. This strategy has since become standard practice in subsequent OpInf studies, as it has been shown to improve numerical stability~\cite{jain2021performance, McQuarrie_2021}.
Note that since the mass fractions fall within the interval [$0,1$], we do not pre-process them. Specifically, $\bS^{(\ell)} = \bQ^{(\ell)} \in \real^{n_x \times K}$ for snapshots of species mass fractions, where $\ell = 6$ corresponds to $\textnormal{O}_2$ mass fraction, $\ell = 7$ to $\textnormal{N}_2$ mass fraction, $\ell = 8$ to $\textnormal{CO}$ mass fraction, $\ell = 9$ to $\textnormal{CO}_2$ mass fraction, and $\ell = 10$ to $\textnormal{H}_2\textnormal{O}$ mass fraction.
After all pre-processing, the original training snapshot $\bQ \in \real^{N \times K}$ is transformed into the scaled snapshot $\bS \in \real^{N \times K}$.

\begin{table}[H]
    \centering
    \caption{Statistics of state variables within the MFiX-PIC training dataset~($t\in[2,14]$s). The solutions on the $22{,}400$ cells over $12{,}000$ snapshots are considered for all $n_v = 10$ state variables.}
    \vspace{-1em}
    \begin{tabular}{r|l l l}
    Variable   & $\phantom{-}$ Minimum   & $\phantom{-}$ Mean   & $\phantom{-}$ Maximum   \\ \hline
    Pressure [$\textnormal{Pa}$] & $\phantom{-} 9.876 \times 10^4$ & $\phantom{-} 1.014 \times 10^5$ & $\phantom{-} 2.600\times10^5$   \\ 
    x velocity [$\textnormal{m/s}$] & $-1.752 \times10$ & $\phantom{-} 2.203 \times 10^{-4}$ & $\phantom{-} 1.742 \times 10$  \\ 
    y velocity [$\textnormal{m/s}$] & $-2.752\times10$ & $-9.470 \times 10^{-1}$ & $\phantom{-} 1.261 \times 10$  \\ 
    z velocity [$\textnormal{m/s}$] & $-1.507\times10$ & $\phantom{-} 1.888\times10^{-4}$ & $\phantom{-} 1.542 \times 10$ \\ 
    Temperature [$\textnormal{K}$] & $\phantom{-} 5.748\times10^2$ & $\phantom{-} 1.133\times10^3$ & $\phantom{-} 2.361\times10^4$ \\ 
    $\textnormal{O}_2$ mass fraction  & $\phantom{-}0$ & $\phantom{-} 1.756\times10^{-1}$ & $\phantom{-} 2.300\times10^{-1}$  \\ 
    $\textnormal{N}_2$ mass fraction & $\phantom{-} 2.166\times10^{-1}$ & $\phantom{-} 7.455\times10^{-1}$ & $\phantom{-} 7.600\times10^{-1}$  \\ 
    $\textnormal{CO}$ mass fraction & $\phantom{-}0$ & $\phantom{-} 8.933\times10^{-4}$ & $\phantom{-} 7.507\times10^{-1}$ \\ 
    $\textnormal{CO}_2$ mass fraction & $\phantom{-}0$ & $\phantom{-} 6.820\times10^{-2}$ & $\phantom{-} 3.799\times10^{-1}$ \\ 
    $\textnormal{H}_2\textnormal{O}$ mass fraction & $\phantom{-} 2.851\times10^{-3}$ & $\phantom{-} 9.810\times10^{-3}$ & $\phantom{-} 1.000\times10^{-2}$
    \end{tabular}
    \label{tab:range_of_variable}
\end{table}

\subsection{Dimensionality reduction} \label{ss:dimensionality_reduction}
Traditionally, OpInf adopts an approach in which snapshots of all state variables are aggregated, and subsequently, a global basis is derived via proper orthogonal decomposition (POD) from the comprehensive training snapshot matrix. 
We thus compute a global POD basis from the transformed snapshot matrix $\bS \in \mathbb{R}^{N \times K}$ of the stacked variables. The reduced singular value decomposition (SVD) of $\bS$, when $N \geq K$, is
\begin{equation}
    \bS = \bV \mathbf{\Sigma} \bW^{\top}
\label{eq:SVD}
\end{equation}
where the columns of $\bV \in \mathbb{R}^{N \times K}$ and  $\bW \in \real^{K \times K}$ are the left and right singular vectors of $\bS$, respectively. The singular values of $\bS$, denoted as $\sigma_1 \geq \sigma_2 \geq \cdots \geq \sigma_K \geq 0$, are on the diagonal of a diagonal matrix $\mathbf{\Sigma} \in \mathbb{R}^{K \times K}$. 
Note that the deterministic SVD of $\bS \in \real^{N \times K}$ has a complexity of \(O(N^2K + K^3)\), making it impractical due to the computational load that grows with the high degree-of-freedom $N$. Thus, we perform randomized SVD via the Python package \texttt{Scikit-learn}~\cite{pedregosa2018scikitlearnmachinelearningpython}. This significantly reduces the complexity to approximately $O(NK\log \omega + \omega^2(N+K))$, where $\omega$ is the target rank and is typically chosen such that $\omega \ll \min(N, K)$~\cite{halko2011finding}.
The $r$ leading columns of the left singular vectors $\bV$ form the rank-$r$ linear POD basis as $\bV_r = [\bv_1, \dots, \bv_r] \in \real^{N \times r}$ whose each column is orthonormal, i.e., $\bV_r^\top \bV_r = \bI_r \in \real^{r \times r}$. We project the scaled snapshot matrix $\bS$ as
\begin{align}
    \widehat{\bS} = \bV_r^{\top}\bS=\left[ {\widehat{\bs}}_0 \hspace{0.7em} \cdots \hspace{0.7em} \widehat{\bs}_{K-1} \right] \in \mathbb{R}^{r \times K}.
\label{eq:low_dim_snapshots}
\end{align}
The full-order scaled state $\bs_k \in \real^{N}$ is approximated as $\bs_k \approx \bV_r \widehat{\bs}_k$. The low-dimensional snapshot matrix $\widehat{\bS}$ contains the data that are used to construct the ROM, which we describe next.

\subsection{Operator Inference with regularization} \label{ss:OpInf_regularization}
OpInf learns the best-fit polynomial ROM from the projected data $\widehat{\bS}$ in Eq.~\eqref{eq:low_dim_snapshots}, potentially with added regularization or constraints. 
For the specific combustion problem considered in~\cite{jain2021performance}, including the cubic term in the right-hand side of the ROM provided a stabilizing effect.
However, forming all possible cubic combinations of the reduced state vector requires $O(r^3)$ operations. Then multiplying the resultant vector by the reduced cubic operator $\widehat{\mathbf{G}} \in \real^{r \times r^3}$ involves $O(r^4)$ multiplications and $O(r(r^3-1))$ additions. Thus, the overall complexity at every time step becomes $O(r^4)$ in an online ROM computation.
Choosing a quadratic polynomial is often a good trade-off between model expressiveness and computational runtime for the ROM simulation~\cite{McQuarrie_2021, Swischuk_2020, qian2022reduced, farcas2023parametric, Zastrow2023, 0x003d183f, Rocha2023}. 
Thus, we opt to learn a quadratic OpInf model as
\begin{equation}
    \frac{\mathrm{d}\widehat{\bs}}{\mathrm{d}t} = \widehat{\bc} + \widehat{\bA} \widehat{\bs} + \widehat{\bH} \left( \widehat{\bs} \otimes \widehat{\bs} \right), \hspace{1.5em} t \in [t_0, t_{K-1}].
\label{eq:polynomial_ROM_eq}
\end{equation}
Here, the operator $\otimes$ indicates a compact Kronecker product that calculates the element-wise multiplication of two vectors while removing redundant quadratic terms. For example, when $\ba = [a_1, a_2, a_3]$, the compact Kronecker product is computed as $\ba \otimes \ba = [a_1^2, a_1a_2, a_1a_3, a_2^2, a_2a_3, a_3^2]$. 
The $\widehat{\bc} \in \real^r$ is a constant term, $\widehat{\bA} \in \real^{r \times r}$ is a linear operator, $\widehat{\bH} \in \mathbb{R}^{r \times r(r+1)/2}$ is a quadratic operator. 
Note that a lifting transformation as in~\cite{qian2022reduced, Swischuk_2020} is not used in this work. This is because directly accessing MFiX-PIC source terms is very involved, and the governing equations feature strong couplings between gas and solid phases, which makes the construction of an explicit lifted representation challenging.
\par

In the model learning process, it is crucial to avoid overfitting the quadratic model to the data due to potential noise or under-resolution of the data, the truncated global POD modes that result in unresolved system dynamics, and the potential for mis-specification of the ROM as fully quadratic, despite the highly nonlinear FOM encapsulated by the governing equations~\eqref{eq:gas_mass_conservation}--\eqref{eq:solid_energy_conservation} and the chemical equations~\eqref{eq:heterogenous reaction}--\eqref{eq:homogeneous reaction}.
Therefore, we adopt Tikhonov regularization for OpInf, as in~\cite{McQuarrie_2021}.
The quadratic OpInf with regularization learns the reduced operators $\widehat{\bc}$, $\widehat{\bA}$, and $\widehat{\bH}$ in the least-squares problem
\begin{align}
    \min_{\widehat{\mathbf{c}}, \widehat{\mathbf{A}}, \widehat{\mathbf{H}}} \quad & 
    \left\{ \sum_{k=0}^{K-1} \left\| \widehat{\mathbf{c}} + \widehat{\mathbf{A}} \widehat{\mathbf{s}}_k + \widehat{\mathbf{H}} \left( \widehat{\mathbf{s}}_k \otimes \widehat{\mathbf{s}}_k \right) - \dot{\widehat{\mathbf{s}}}_k \right\|_2^2 + \lambda_1 \left( \left\| \widehat{\mathbf{c}} \right\|_2^2 + \left\| \widehat{\mathbf{A}} \right\|_F^2 \right) + \lambda_2 \left\| \widehat{\mathbf{H}} \right\|_F^2 \right\} ,
\label{eq:optimization_regularized}
\end{align}
where $\widehat{\bs}_k \equiv \widehat{\bs}(t_k) $ represents the reduced state at the time step $k$. Note that the Kronecker product introduces scaling differences between the entries of the quadratic operator $\widehat{\bH}$, and those in the constant operator $\widehat{\bc}$ and the linear operator $\widehat{\bA}$, when subjected to a single regularization parameter.
This motivates us to use two regularization parameters $\lambda_1 > 0$ and $\lambda_2 > 0$, such that $\lambda_1$ penalizes $\widehat{\bc}$ and $\widehat{\bA}$, and $\lambda_2$ penalizes $\widehat{\bH}$.
Here, $\dot{\widehat{\bs}}_k \equiv \dot{\widehat{\bs}}(t_k)$ denotes the time derivative of the reduced state at the time step $k$, and it can be computed using a suitable time derivative approximation scheme. We use a $6$th-order central finite difference method, with biased forward and backward differencing applied at the initial and final time steps.
Equation~\eqref{eq:optimization_regularized} can be written compactly as
\begin{align}
    \min_{\bO} \left\| \mathbf{DO}^{\top} - \dot{\widehat{\bS}}^{\top} \right\|_{F}^{2} + \left\|\mathbf{\Lambda} \bO^{\top} \right\|_{F}^{2}, \quad \text{where} \quad
    \bO = \left[\widehat{\bc} \hspace{0.7em} \widehat{\bA} \hspace{0.7em} \widehat{\bH} \right], \hspace{0.7em}
    \bD = \left[\mathbf{1}_K \hspace{0.5em} \widehat{\bS}^{\top} \hspace{0.5em} \left(\widehat{\bS} \otimes \widehat{\bS} \right)^{\top} \right],
\label{eq:regularization_matrix_form}
\end{align}
and where $\bO \in \mathbb{R}^{r \times d(r)}$ and $\bD \in \mathbb{R}^{K \times d(r)}$.
Here, $\dot{\widehat{\bS}} \in \real^{r \times K}$ is a matrix whose columns are $\dot{\widehat{\bs}}_k$, and $\mathbf{1}_K \in \mathbb{R}^K$ is a column vector of length $K$ with all entries set to one. Also, $\mathbf{\Lambda} = \textnormal{diag}(\lambda_1, \lambda_1 \bI_{r}, \lambda_2 \bI_{r(r+1)/2}) \in \mathbb{R}^{d(r) \times d(r)}$ serves as a diagonal regularization matrix, where $\bI_{r}$ and $\bI_{r(r+1)/2}$ are identity matrices with dimensions $r$ and $r(r+1)/2$, respectively, making $d(r) = 1+r+r(r+1)/2$. In practice, we must carefully choose the regularization parameters $\lambda_1$ and $\lambda_2$; see~\Cref{ss:regularization_hyperparameters_selection} for details. 
The reduced operators inferred by solving Eq.~\eqref{eq:regularization_matrix_form} provide the best fits of reduced state trajectories of the ROM Eq.~\eqref{eq:polynomial_ROM_eq} within the training regime in a minimum residual sense.

\subsection{State constraints to enforce physical consistency of the operator inference reduced-order model} \label{ss:species_limiter}
Low-dimensional surrogate modeling is challenging for nonlinear multiscale and multiphysics problems. Consequently, there have been many research advances to ensure better predictive capabilities of these models.
As discussed in~\Cref{sec:intro}, various structure-preserving methods offer an approach to enforce physical consistency in OpInf ROMs. While these approaches have successfully addressed issues of mathematical stability, energy preservation, and conservation of mechanistic principles, they are not guaranteed to provide ROM predictions that are consistent with the physical constraints inherent in the FOM. In this section, we propose a method of maintaining these intrinsic physical constraints within the state prediction via OpInf ROMs.

\subsubsection{Limitation of the standard operator inference reduced-order model} \label{sss:Outline_state_constrained_OpInf}
The standard OpInf propagates states in the low-dimensional space from the initial reduced state $\widehat{\bs}_0$ to the final reduced state $\widehat{\bs}_{K_f-1}$ through a numerical time integration. Here, $K_f$ represents the total number of snapshots forecasted through OpInf beyond the training regime. This is illustrated in the dashed arrows in~\Cref{fig:species_limiter_step}, which highlights the contrast in state propagation when using standard OpInf compared to the proposed scheme with state constraints. Once the temporal evolution of all the reduced states $\widehat{\bs}_k$ is done, the high-dimensional states are reconstructed by $\widetilde{\bS} = \bV_r \widehat{\bS} \in \real^{N \times K_f}$. 
However, the reconstructed high-dimensional state $\widetilde{\bs}_k$ is not guaranteed to comply with the potential constraints posed by the physics of the FOM.
Particularly for the char combustion data considered herein, some predicted species mass fractions, which are in $\widetilde{\bS}^{(\ell)}$, where $\ell=6,7,8,9,10$, see~\Cref{ss:data_pre_processing}, may not fall within the range $[0,1]$. In fact, even the reconstruction of the initial reduced state, which is $\bV_r \widehat{\bs}_0$, is not assured to provide mass fractions consistent with such a constraint.

\subsubsection{Design of state constraints}     \label{sss:design_state_constraints}
In the proposed framework of OpInf ROM with state constraints, we enforce the state constraints in the reconstructed high-dimensional state $\widetilde{\bs}_k \in \real^{N}$ at every time step during the state propagation. Since the available FOM data of the char combustion does not provide sufficient information to determine whether pressure, temperature, and velocities inherit physical constraints, we limit our state constraints to the species mass fraction variables. This is problem-specific; thus, if physical limitations regarding other variables are available from prior knowledge of the FOM, different constraints may have to be applied.
By embedding state constraints for the species mass fraction variables, we aim to enhance the physical consistency of the predictions made by the OpInf ROM, thereby improving its ability to stably and accurately capture the complex dynamics of char combustion processes while maintaining computational efficiency. \par

\begin{figure}[htbp]
    \centering
    \begin{tikzpicture}[
        node distance=1.0cm and 1.0cm,
        every node/.style={align=center},
        lowdim/.style={font=\small},
        highdim/.style={font=\small},
        vec/.style={->, thick, C02},
        sl/.style={->, thick, C02},
        transform/.style={->, thick, C02},
        reduction/.style={->, thick, C02},
    ]
    
    \node[highdim] (S00) {${\bs}_0$};
    \node[highdim, right=of S00] (S0) {$\widetilde{\bs}_0$};
    \node[highdim, right=of S0] (S0*) {$\widetilde{\bs}_0^*$};
    \node[highdim, right=of S0*] (S1) {$\widetilde{\bs}_1$};
    \node[highdim, right=of S1] (S1*) {$\widetilde{\bs}_1^*$};
    \node[highdim, right=of S1*] (S2) {$\widetilde{\bs}_2$};
    \node[highdim, right=of S2] (S2*) {$\widetilde{\bs}_2^*$};
    
    \node[lowdim, below=of S0] (s0) {$\widehat{\bs}_0$};
    \node[lowdim, right=of s0] (s0*) {$\widehat{\bs}_0^{*}$};
    \node[lowdim, right=of s0*] (s1) {$\widehat{\bs}_1$};
    \node[lowdim, right=of s1] (s1*) {$\widehat{\bs}_1^{*}$};
    \node[lowdim, right=of s1*] (s2) {$\widehat{\bs}_2$};
    \node[lowdim, right=of s2] (s2*) {$\widehat{\bs}_2^{*}$};

    \node[lowdim, right=of s2*] (s3) {$\widehat{\bs}_3$};
    \node[lowdim, right=0.1cm of s3] (dots_low) {$\cdots$};

    \draw[vec] (s0*) -- node[midway,below] {\textcolor{C02}{ROM}} (s1);
    \draw[vec] (s1*) -- node[midway,below] {\textcolor{C02}{ROM}} (s2);
    \draw[vec] (s2*) -- node[midway,below] {\textcolor{C02}{ROM}} (s3);

    \draw[C01, thick, dashed, ->, bend left=-65] (s0) to node[midway,below,sloped] {ROM} (s1);
    \draw[C01, thick, dashed, ->, bend left=-65] (s1) to node[midway,below,sloped] {ROM} (s2);
    \draw[C01, thick, dashed, ->, bend left=-65] (s2) to node[midway,below,sloped] {ROM} (s3);

    \draw[sl] (S0) -- node[midway,above] {species \\ limiters} (S0*);
    \draw[sl] (S1) -- node[midway,above] {species \\ limiters} (S1*);
    \draw[sl] (S2) -- node[midway,above] {species \\ limiters} (S2*);
    
    \draw[transform] (s0) -- node[midway,right] {$\bV_r \times$} (S0);
    \draw[transform] (s1) -- node[midway,right] {$\bV_r \times$} (S1);
    \draw[transform] (s2) -- node[midway,right] {$\bV_r \times$} (S2);
    
    \draw[reduction] (S0*) -- node[midway,right] {$\bV_r^{\top} \times$} (s0*);
    \draw[reduction] (S1*) -- node[midway,right] {$\bV_r^{\top} \times$} (s1*);
    \draw[reduction] (S2*) -- node[midway,right] {$\bV_r^{\top} \times$} (s2*);
    
    \node[left=of S00, xshift=0.6cm] {\\ High-dim. \\ space};
    \node[left=of s0, xshift=-1cm] {\\ Low-dim. \\ space};
    \draw[->, thick, dotted, C02] (S00) -- node[midway,below, xshift=-10pt, yshift=3pt] {$\bV_r^\top \times$} (s0);

    \usetikzlibrary{calc}
    \usetikzlibrary{backgrounds}
    \pgfdeclarelayer{background}
    \pgfsetlayers{background,main}
    \begin{scope}[on background layer]
        \draw[solid, thick, black]
            ($(S0.north west)+(-4.4cm, 1.0cm)$) rectangle 
            ($(s2*.south east)+(2.7cm,-1.5cm)$);
    \end{scope}
    
    \end{tikzpicture}
    \caption{Schematic of the state propagation in the ROM with species limiters, which are state constraints for the species mass fractions. The dashed arrows represent the reduced state propagation via the standard OpInf ROM, and the solid arrows represent the state propagation via the OpInf with species limiters. The orthogonal basis $\bV_r \in \real^{N \times r}$ is used for projection between the low-dimensional space $\real^{r}$ and the high-dimensional space $\real^{N}$.}
    \label{fig:species_limiter_step}
\end{figure}
To design state constraints on species mass fractions, we use the knowledge of the variable ranges as well as boundary conditions of the FOM, as shown in~\Cref{tab:combustion_parameters}, and chemical reactions~\eqref{eq:heterogenous reaction}--\eqref{eq:homogeneous reaction}. 
Concerning the air inflow boundary conditions of the combustion model, the maximum permissible values of the mass fractions of $\textnormal{O}_2$, $\textnormal{N}_2$, and $\textnormal{H}_2\textnormal{O}$ are dictated by the corresponding constituents within the air inflow, given that these components are not produced from both Eqs.~\eqref{eq:heterogenous reaction}--\eqref{eq:homogeneous reaction}.
We denote those upper bounds as $Y^{U}$. Those for $\textnormal{O}_2$, $\textnormal{N}_2$, and $\textnormal{H}_2\textnormal{O}$ are $Y_{\textnormal{O}_2}^{U}=0.23$, $Y_{\textnormal{N}_2}^{U}=0.76$, and $Y_{\textnormal{H}_2\textnormal{O}}^{U}=0.01$, see~\Cref{tab:range_of_variable}.
However, the $\textnormal{CO}$ and $\textnormal{CO}_2$ do not constitute the inlet air inflow and are the products of the two chemical reactions considered. Thus, we set $Y_{\textnormal{CO}}^{U}$ and $Y_{\textnormal{CO}_2}^{U}$ to $1$, the possible maximum value of species mass fractions.
Moreover, mass fractions in the FOM data, and in general, are always non-negative, see~\Cref{tab:range_of_variable} for the range of FOM states in the training regime.
Thus, we enforce a lower bound of zero for all species mass fractions ($Y_{\textnormal{O}_2}^{L} = Y_{\textnormal{N}_2}^{L} = Y_{\textnormal{CO}}^{L} = Y_{\textnormal{CO}_2}^{L} = Y_{\textnormal{H}_2\textnormal{O}}^{L} = 0$) throughout both the training and testing regimes.
In sum, we formulate the state constraints of all species mass fractions as
\begin{align} 
    \label{eq:species_limiter_O2}
    Y_{\textnormal{O}_2} &\leftarrow \max \left( \min \left( Y_{\textnormal{O}_2}, Y_{\textnormal{O}_2}^{U} \right), Y_{\textnormal{O}_2}^{L} \right), \\
    \label{eq:species_limiter_N2}
    Y_{\textnormal{N}_2} &\leftarrow \max \left( \min \left( Y_{\textnormal{N}_2}, Y_{\textnormal{N}_2}^{L} \right), Y_{\textnormal{N}_2}^{L} \right), \\
    \label{eq:species_limiter_CO}
    Y_{\textnormal{CO}} &\leftarrow \max \left( \min \left( Y_{\textnormal{CO}}, Y_{\textnormal{CO}}^{U} \right), Y_{\textnormal{CO}}^{L} \right), \\
    \label{eq:species_limiter_CO2}
    Y_{\textnormal{CO}_2} &\leftarrow \max \left( \min \left( Y_{\textnormal{CO}_2}, Y_{\textnormal{CO}_2}^{U} \right), Y_{\textnormal{CO}_2}^{L} \right), \\
    \label{eq:species_limiter_H2O}
    Y_{\textnormal{H}_2\textnormal{O}} &\leftarrow \max \left( \min \left( Y_{\textnormal{H}_2\textnormal{O}}, Y_{\textnormal{H}_2\textnormal{O}}^{U} \right), Y_{\textnormal{H}_2\textnormal{O}}^{L} \right).
\end{align}
\par

\subsubsection{ROM state propagation with state constraints} \label{sss:state_propagation_state_constraints}
In~\Cref{fig:species_limiter_step}, we illustrate that states propagate across two spaces within the OpInf ROM with species limiters. We start from the true scaled high-dimensional state~$\bs_0$ and project it to the initial reduced state~$\widehat{\bs}_0$.
To maintain the physical consistency from the initial time step, the initial reduced state $\widehat{\bs}_0$ first must be lifted to the high-dimensional state $\widetilde{\bs}_0$ that contains all variables.
The state constraints, formulated as Eqs.~\eqref{eq:species_limiter_O2}--\eqref{eq:species_limiter_H2O}, are directly enforced to $\widetilde{\bs}_0$, leading to the modified high-dimensional state $\widetilde{\bs}_0^*$. This subsequently needs to be projected back to $\widehat{\bs}^*_0$, so that the ROM propagates $\widehat{\bs}_0^*$ to $\widehat{\bs}_1$. In this manner, we propagate states over time within the low-dimensional space, thereby maintaining computational efficiency. This iterative procedure continues until the final time.
Once the entire temporal propagation is done, we unscale the modified high-dimensional states $\widetilde{\bs}^*_k$ ($k=0,1,\ldots,K_f-1$), which contain scaled variables such as pressure, temperature, and velocities, via the inversion of Eqs.~\eqref{eq:shifting}--\eqref{eq:scaling}, obtaining the high-dimensional states in the original scale $\widetilde{\bq}^*_k$.
Note that since we only constrain the species mass fraction variables, which are not scaled initially, we obviate the necessity to restore the original scales of $\widetilde{\bs}_k$ at every time step during the state propagation. Nevertheless, should limiting conditions pertain to pre-processed variables, restoring their original scales in the high-dimensional space becomes necessary at each time step prior to enforcing state constraints.
\par
The presented approach departs from the conventional offline-online decomposition in model reduction by incorporating online evaluations that partially reintroduce FOM information through state corrections. While the reduced dynamics evolve efficiently via a ROM, the correction step can be interpreted as providing a \textit{closure mechanism} that enforces physical constraints the ROM alone may violate. This establishes a hybrid simulation:~the ROM captures and evolves dominant dynamics efficiently, and the FOM-informed correction ensures physicality.
~\Cref{al:OpInf_with_SL} provides the algorithmic process of OpInf with state constraints. Note that the model learning process, including data pre-processing, SVD, and regularization, requires $K$ training snapshots, and the temporal propagation with state constraints can go beyond this training regime, to $K_f$ time steps.
\begin{algorithm}
\caption{Operator Inference with state constraints} \label{al:OpInf_with_SL}
\begin{algorithmic}[1]
    \Procedure{OpInf with state constraints}{}
    
    \State Transform original snapshots $\bQ \in \real^{N \times K}$ into $\bS \in \real^{N \times K}$ as in Eqs.~\eqref{eq:shifting}--\eqref{eq:scaling}
    
    \State Compute randomized SVD with a target rank $r_t$:
    $\bS = \bV_{r_t} \mathbf{\Sigma}_{r_t} \bW_{r_t}^{\top}$ where $\bV_{r_t} \in \real^{N \times r_t}$, $\mathbf{\Sigma}_{r_t} \in \real^{{r_t} \times {r_t}}$, and $\bW_{r_t} \in \real^{K \times {r_t}}$

    \State Select a rank-$r$ low-dimensional basis $\bV_r = \bV_{r_t}[:,:r]$
    
    \State Project $\bS$ to the linear low-dimensional subspace: $\widehat{\bS} = \bV_r^{\top} \bS \in \real^{r \times K}$

    \State Evaluate the time derivative of the reduced snapshot $\widehat{\bS}$

    \Procedure{Regularization}{}
    
    \State Search hyperparameters $\lambda_1^{*}$ and $\lambda_2^{*}$ based on the key performance indicator criterion as in Eq.~\eqref{eq:thermal_energy_eq}
    
    \EndProcedure
    
    \Procedure{Time propagation over $K_f$ time steps with state constraints}{}
    \For{$k=0, \dots, K_f-1$}
        \State Reconstruct the high-dimensional state: $\widetilde{\bs}_{k} = \bV_r \widehat{\bs}_k \in \real^{N}$
        
        \State Apply the state constraints Eqs.~\eqref{eq:species_limiter_O2}--\eqref{eq:species_limiter_H2O}, to $\widetilde{\bs}_k$, obtaining a modified full-state $\widetilde{\bs}_k^*$

        \State Project $\widetilde{\bs}_k^{*}$ to the low-dimensional subspace: $\widehat{\bs}_k^{*} = \bV_r^{\top} \widetilde{\bs}_k^{*}$

        \State Propagate from $\widehat{\bs}_k^{*}$ to $\widehat{\bs}_{k+1}$ by time integration of Eq.~\eqref{eq:polynomial_ROM_eq}

    \EndFor
    \EndProcedure

    \State Restore high-dimensional states $\widetilde{\bq}_{k}^{*}$ in the original scales from $\widetilde{\bs}_{k}^{*}$
    
    \State \Return $\bQ^{*} = [ \widetilde{\bq}_0^{*} \hspace{1em} \widetilde{\bq}_1^{*} \hspace{1em} \cdots \hspace{1em} \widetilde{\bq}_{K_f-1}^{*} ] \in \real^{N \times K_f}$
\EndProcedure
\end{algorithmic}
\end{algorithm}

\subsubsection{Stabilizing effects of ROM with state constraints and a global basis} \label{sss:stabilizing_effect}
The species limiters that act on a part of the state have a stabilizing effect on the entire state vector. This stabilizing effect is further reinforced by the global POD basis, which plays a key role in enhancing overall stability.~\Cref{fig:state_correction_illustration} provides a graphical illustration of the state propagation, embedded with state constraints.
\begin{figure}[htbp]
\centering
\begin{tikzpicture}[
    box/.style={draw, minimum width=1.2cm, minimum height=1.5cm, inner sep=0pt},
    box2/.style={draw, minimum width=1.5cm, minimum height=1.0cm, inner sep=0pt},
    short vec/.style={draw, minimum width=0.4cm, minimum height=1.5cm, inner sep=0pt},
    long vec/.style={draw, minimum width=0.5cm, minimum height=6cm, inner sep=0pt},
    short var/.style={draw, minimum width=0.5cm, minimum height=1.0cm, inner sep=0pt},
    hatched long box/.style={draw, pattern=north east lines, minimum width=0.5cm, minimum height=3.0cm, inner sep=0pt},
    hatched short box/.style={draw, pattern=north east lines, minimum width=0.4cm, minimum height=1.5cm, inner sep=0pt}
]
    \draw (-13, 7.2) node[long vec] {};
    \node at (-13, 10.5) {$\widetilde{\bs}_k$};

    \draw (-13, 9.7) node[short var] {};
    \draw (-13, 8.7) node[short var] {};
    \draw (-13, 7.7) node[short var] {};
    
    \draw (-13, 6.7) node[short var] {};
    \draw (-13, 5.7) node[short var] {};
    \draw (-13, 4.7) node[short var] {};

    \draw[decorate,decoration={brace}] (-12.7, 10.2) -- (-12.7, 9.2);
    \node at (-11.9, 9.7) {$\widetilde{\bp}^{\textnormal{scaled}}_k$};
    \node[rotate=90] at (-11.9, 8.7) {$\cdots$};
    \draw[decorate,decoration={brace}] (-12.7, 8.2) -- (-12.7, 7.2);
    \node at (-11.9, 7.7) {$\widetilde{\bT}^{\textnormal{scaled}}_k$};
    \draw[decorate,decoration={brace}] (-12.7, 7.2) -- (-12.7, 6.2);
    \node at (-11.9, 6.7) {$\widetilde{\bY}_{\textnormal{O}_2,k}$};
    \node[rotate=90] at (-11.9, 5.7) {$\cdots$};
    \draw[decorate,decoration={brace}] (-12.7, 5.2) -- (-12.7, 4.2);
    \node at (-11.9, 4.7) {$\widetilde{\bY}_{\textnormal{H}_2\textnormal{O},k}$};
    
    \node at (-10.5, 7.2) {$\longrightarrow$};

    \node at (-10.5, 6.8) {Apply};
    \node at (-10.5, 6.3) {species};
    \node at (-10.5, 5.8) {limiters};

    \draw (-9.25, 7.2) node[long vec] {};
    \node at (-9.25, 10.5) {$\widetilde{\bs}_k^*$};

    \usetikzlibrary{patterns}
    \draw (-9.25, 5.7) node[hatched long box] {};

    \draw[dashed] (-8.5, 4) -- (-8.5, 12);

    \node at (-11, 11.5) {\text{Application of species limiters}};

    \foreach \x in {1.2, 2.4, 3.6, 4.8, 6.0, 7.2} {
        \draw (-8.50+\x, 9.45) node[box] {};
    }
    \node at (-7.3, 9.45) {$\bV_{p_g}^\top$};
    \node at (-6.1, 9.45) {$\cdots$};
    \node at (-4.9, 9.45) {$\bV_{T_g}^\top$};
    \node at (-3.7, 9.45) {$\bV_{Y_{\textnormal{O}_2}}^\top$};
    \node at (-2.5, 9.45) {$\cdots$};
    \node at (-1.3, 9.45) {$\bV_{Y_{\textnormal{H}_2\textnormal{O}}}^\top$};

    \node at (-4.25, 10.75) {$\bV_r^\top$};
    \draw[decorate,decoration={brace}] (-7.90, 10.35) -- (-0.75, 10.35);

    \draw (-0.25, 7.2) node[long vec] {};
    \node at (-0.25, 10.5) {$\widetilde{\bs}_k^*$};

    \usetikzlibrary{patterns}
    \draw (-0.25, 5.7) node[hatched long box] {};
    
    \node at (0.5, 9.45) {=};

    \draw (1.2, 9.45) node[hatched short box] {};
    \node at (1.2, 10.5) {$\widehat{\bs}_k^*$};

    \node at (-3.3, 11.5) {\text{Projection of corrected high-dimensional state}};
\end{tikzpicture}
\caption{Interpretation of the state constraints-embedded state propagation, as presented in ~\Cref{fig:species_limiter_step}. The process is illustrated for a single time step. The left-hand side of the vertical dotted line represents the enforcement of species limiters, and the shaded area of $\widetilde{\bs}_k^*$ represents the corrected values. The right-hand side of the vertical dotted line represents the projection of the partially corrected high-dimensional state, modifying all the entries of the reduced state $\widehat{\bs}_k^*$.}
\label{fig:state_correction_illustration}
\end{figure}
Since we use the global POD basis computed from stacked state snapshots, see \Cref{ss:dimensionality_reduction}, the linear combination of the $r$ columns of $\bV_{r}[(\ell-1)n_x:\ell n_x - 1, :] \in \real^{n_x \times r}$ and the $r$ entries of $\widetilde{\bs}_k$ reconstructs each $\ell$th state vector, which is denoted as $\widetilde{\bs}_k[(\ell-1)n_x:\ell n_x - 1]$. For example, the reconstructed scaled pressure vector at time step $k$, $\widetilde{\bp}_k^{\textnormal{scaled}} \in \real^{n_x}$, is computed as a linear combination of the columns of $\bV_{p_g}$ and the entries of $\widehat{\bs}_k$, where $\bV_{p_g}$ is the first $n_x$ rows of the global basis $\bV_r \in \real^{N \times r}$. In other words, when $\bV_{p_g} = \left[ \bv_{p_g,1}, \cdots, \bv_{p_g,r} \right] \in \real^{n_x \times r}$, $\widetilde{\bp}_k^{\textnormal{scaled}}$ is computed as $\sum_{\eta=1}^{r} \widehat{\bs}_{k,\eta} \cdot \bv_{p_g,\eta}$.
After all the states are reconstructed, leading to $\widetilde{\bs}_k$, we enforce the state constraints to the mass fraction states. Thus, $\widetilde{\bY}_{\textnormal{O}_2, k}$, $\widetilde{\bY}_{\textnormal{N}_2, k}$, $\widetilde{\bY}_{\textnormal{CO}, k}$, $\widetilde{\bY}_{\textnormal{CO}_2, k}$, and $\widetilde{\bY}_{\textnormal{H}_2\textnormal{O}, k}$ are corrected, producing a modified high-dimensional state $\widetilde{\bs}_k^*$. In \Cref{fig:state_correction_illustration}, we indicate the modified states with a shaded area.
Afterward, we compress $\widetilde{\bs}_k^*$ through the global POD basis $\bV_r$, resulting in the low-dimensional state $\widehat{\bs}_k^*$. We can view the numerical aspect of this projection in two ways. First, it is a linear combination of all the columns of $\bV_r^\top$ and all the entries of $\widetilde{\bs}_k^*$, meaning $\widehat{\bs}_k^* = \sum_{n=1}^{N} \widetilde{\bs}^{*}_{k,n} \cdot \bV_r^\top[:,n]$.
Another is that each entry of $\widehat{\bs}_k^*$, denoted as $\widehat{\bs}^{*}_{k,\eta}$ ($\eta = 1,2,\cdots,r$), is the outer product between a row vector of $\bV_r^\top$, which is $\bV_r^\top[\eta,:]$, and a column vector $\widetilde{\bs}^*_{k}$. Either way, all the entries of the resultant low-dimensional state $\widehat{\bs}_k^*$ are modified. Because the ROM advances the reduced states as a coupled system, these changes propagate to all variables, providing an additional source of stabilization. Consequently, in the next time step $k+1$, even the reconstructed high-dimensional states not directly constrained by species limiters are affected. Thus, the proposed approach benefits from two complementary stabilizing mechanisms: (i) the state constraints when using the global POD basis, and (ii) the inherent coupling of the ROM dynamics. 
In our case, these combined effects stabilize the evolution of all state variables, as further demonstrated in the numerical results in \Cref{sss:results}.

\subsubsection{ROM prediction with state constraints and a block-structured basis}    \label{sss:block_POD_basis}
\par
Constructing an individual basis for each variable, rather than computing a stacked global POD basis from the entire training snapshot matrix $\bS \in \real^{N \times K}$, as illustrated in the previous section, leads to more interpretable ROM coefficients. This approach is discussed in detail in~\cite[Sec.~4.3.2]{benner2020operator2}, where the authors used a block-diagonal projection matrix to address the topological structure of a highly coupled nonlinear problem.
Specifically in our case, we can compute a separate basis for the training snapshot matrix of the $\ell$th state variable, $\bS^{(\ell)} \in \real^{n_x \times K}$ (see Eq.~\eqref{eq:scaling}). When $n_x \geq K$, the reduced SVD of $\bS^{(\ell)}$ is given by
$$\bS^{(\ell)} = \bV^{(\ell)} \mathbf{\Sigma}^{(\ell)} \bW^{(\ell) \top},$$
where $\bV^{(\ell)} \in \real^{n_x \times K}$, $\mathbf{\Sigma}^{(\ell)} \in \real^{K \times K}$, and $\bW^{(\ell)} \in \real^{K \times K}$.
The linear POD basis for the $\ell$th state variable is obtained as the leading $r^{(\ell)}$ columns of the left singular vector matrix $\bV^{(\ell)}$, denoted as $\bV_r^{(\ell)} \in \real^{n_x \times r^{(\ell)}}$.
Each reduction dimension $r^{(\ell)}$ is chosen separately for each state.
\par
After computing the individual bases for all state variables, each basis is used to project its corresponding high-dimensional state into a reduced state. To clarify, referring to Eq.~\eqref{eq:solution_variables}, we define the bases for all variables as follows: $\bV_{\bp,r} = \bV_r^{(\ell=1)}$ for pressure, $\bV_{\bv_x,r} = \bV_r^{(\ell=2)}$ for $x$-velocity, $\cdots$, $\bV_{\bY_{\textnormal{H}_2\textnormal{O}}, r} = \bV_r^{(\ell=10)}$ for $\textnormal{H}_2\textnormal{O}$ mass fraction. Using these bases, the high-dimensional states are projected into their reduced states: $\widehat{\bs}_{p,k} = \bV_{\bp,r}^\top \bp^{\textnormal{scaled}}_k \in \real^{r^{(1)}}$ for pressure, $\widehat{\bs}_{v_x,k} = \bV_{\bv_x, r}^\top \bv_{x,k}^{\textnormal{scaled}} \in \real^{r^{(2)}}$ for $x$-velocity, $\cdots$, $\widehat{\bs}_{Y_{\textnormal{H}_2\textnormal{O}}, k} = \bV_{\bY_{\textnormal{H}_2\textnormal{O},r}}^\top \bY_{\textnormal{H}_2\textnormal{O}, k} \in \real^{r^{(10)}}$ for $\textnormal{H}_2\textnormal{O}$ mass fraction. Since each variable is projected differently, we construct a block-structured basis
\begin{align*}
    \bV_r = \textnormal{diag} \left( \bV_{\bp,r}, \bV_{\bv_x,r}, \cdots, \bV_{\bY_{\textnormal{H}_2\textnormal{O}}, r} \right) \in \real^{N \times r} ,
\end{align*}
where $r = \sum_{\ell=1}^{10} r^{(\ell)}$. This basis projects the entire full state $\bs_k$ into the reduced state $\widehat{\bs}_k \in \real^{r}$ via $\widehat{\bs}_k = \bV_r^\top \bs_k$. 
The variables remain decoupled when reconstructing the full state $\widetilde{\bs}_k$ from the entire reduced state $\widehat{\bs}_k$ through the block-structured basis $\bV_r$ as
\begin{equation}
\widetilde{\bs}_k = 
\begin{bmatrix}
\widetilde{\bp}^{\textnormal{scaled}}_k \\
\widetilde{\bv}_{x,k}^{\textnormal{scaled}} \\
\vdots \\
\widetilde{\bY}_{\textnormal{H}_2 \textnormal{O}, k}
\end{bmatrix}
\approx
\begin{bmatrix}
\bV_{\bp,r} &  &  & \\
 & \bV_{\bv_x, r} &  & \\
 &  & \ddots &  \\
 &  &  & \bV_{\bY_{\textnormal{H}_2 \textnormal{O}}, r} 
\end{bmatrix}
\begin{bmatrix}
\widehat{\bs}_{p,k} \\
\widehat{\bs}_{v_x,k} \\
\vdots \\
\widehat{\bs}_{Y_{\textnormal{H}_2 \textnormal{O}}, k}
\end{bmatrix}.
\label{eq:block_POD_decoupling}
\end{equation}
\par
We next discuss the effect of using a block-stuctured projection matrix in the context of state constraints.~\Cref{fig:block_structured_POD} illustrates the projection of the corrected high-dimensional states $\widetilde{\bs}_k^*$ using the block-structured POD basis, resulting in the modified reduced states $\widehat{\bs}_k^*$. Similar to~\Cref{fig:state_correction_illustration}, the shaded areas in $\widetilde{\bs}_k^*$ represent the parts corrected by the state constraints on mass fractions.
Because of the block structure of $\bV_r^\top$, the coefficients in the reduced states $\widehat{\bs}_k^*$ associated with variables other than species mass fractions remain unchanged. In particular, the first $\sum_{\ell=1}^5 r^{(\ell)}$ entries of $\widehat{\bs}_k^*$ (corresponding to unmodified states such as pressure, temperature, velocities in our case) are unaffected by the state constraints.
However, once the ROM evolves the \emph{partially modified} reduced state $\widehat{\bs}_k^*$ in time, its coupled dynamics cause all state variables to be indirectly affected. Consequently, similar to the case of the global basis, at the next time step $k+1$, the components of the reduced state $\widehat{\bs}_k^*$ for pressure, temperature, and velocities differ from those that would have been obtained without mass-fraction correction at time step $k$.
In other words, with a block-structured basis, state constraints still provide a stabilizing effect on all variables through the coupled ROM dynamics, though not as strongly as in the global basis case, which has two stabilizing mechanisms, as discussed in~\Cref{sss:stabilizing_effect}.
\begin{figure}[htbp]
\centering
\begin{tikzpicture}[
    box/.style={draw, minimum width=1.2cm, minimum height=1.5cm, inner sep=0pt},
    box2/.style={draw, minimum width=1.5cm, minimum height=1.0cm, inner sep=0pt},
    short vec1/.style={draw, minimum width=0.4cm, minimum height=1.5cm, inner sep=0pt},
    long vec1/.style={draw, minimum width=0.5cm, minimum height=6cm, inner sep=0pt},
    short var/.style={draw, minimum width=0.5cm, minimum height=1.0cm, inner sep=0pt},
    hatched long box/.style={draw, pattern=north east lines, minimum width=0.5cm, minimum height=3.0cm, inner sep=0pt},
    hatched short box/.style={draw, pattern=north east lines, minimum width=0.4cm, minimum height=1.5cm, inner sep=0pt},
    wide box/.style={draw,  minimum width=1.5cm, minimum height=0.7cm, inner sep=0pt},
    wide big box/.style={draw, minimum width=6.95cm, minimum height=4cm, inner sep=0pt},
    long vec2/.style={draw, minimum width=0.5cm, minimum height=6.95cm, inner sep=0pt},
    shaded vec/.style={draw, pattern=north east lines, minimum width=0.5cm, minimum height=3.475cm, inner sep=0pt},
    short vec2/.style={draw, minimum width=0.5cm, minimum height=4cm, inner sep=0pt},
    short shaded vec/.style={draw, pattern=north east lines, minimum width=0.5cm, minimum height=2cm, inner sep=0pt}
]
    \draw (-13, 7.2) node[long vec1] {};
    \node at (-13, 10.5) {$\widetilde{\bs}_k$};

    \draw (-13, 9.7) node[short var] {};
    \draw (-13, 8.7) node[short var] {};
    \draw (-13, 7.7) node[short var] {};
    
    \draw (-13, 6.7) node[short var] {};
    \draw (-13, 5.7) node[short var] {};
    \draw (-13, 4.7) node[short var] {};

    \draw[decorate,decoration={brace}] (-12.7, 10.2) -- (-12.7, 9.2);
    \node at (-11.9, 9.7) {$\widetilde{\bp}^{\textnormal{scaled}}_k$};
    \node[rotate=90] at (-11.9, 8.7) {$\cdots$};
    \draw[decorate,decoration={brace}] (-12.7, 8.2) -- (-12.7, 7.2);
    \node at (-11.9, 7.7) {$\widetilde{\bT}^{\textnormal{scaled}}_k$};
    \draw[decorate,decoration={brace}] (-12.7, 7.2) -- (-12.7, 6.2);
    \node at (-11.9, 6.7) {$\widetilde{\bY}_{\textnormal{O}_2,k}$};
    \node[rotate=90] at (-11.9, 5.7) {$\cdots$};
    \draw[decorate,decoration={brace}] (-12.7, 5.2) -- (-12.7, 4.2);
    \node at (-11.9, 4.7) {$\widetilde{\bY}_{\textnormal{H}_2\textnormal{O},k}$};
    
    \node at (-10.5, 7.2) {$\longrightarrow$};

    \node at (-10.5, 6.8) {Apply};
    \node at (-10.5, 6.3) {species};
    \node at (-10.5, 5.8) {limiters};

    \draw (-9.25, 7.2) node[long vec1] {};
    \node at (-9.25, 10.5) {$\widetilde{\bs}_k^*$};

    \usetikzlibrary{patterns}
    \draw (-9.25, 5.7) node[hatched long box] {};

    \draw[dashed] (-8.5, 4) -- (-8.5, 12);

    \node at (-11, 11.5) {\text{Application of species limiters}};

    \draw (-4.5, 8.2) node[wide big box] {};

    \draw (-2.725-4.5, 5.65+4.2) node[wide box] {};
    \node at (-2.725-4.5, 5.65+4.2) {$\bV_{\bp, r}^{\top}$};
    \node[rotate=135] at (-1.725-4.5, 5+4.2) {$\cdots$};

    \draw (-0.725-4.5, 4.3+4.2) node[wide box] {};
    \node at (-0.725-4.5, 4.3+4.2) {$\bV_{\bT, r}^{\top}$};

    \draw (0.85-4.57, 3.60+4.2) node[wide box] {};
    \node at (0.85-4.57, 3.60+4.2) {$\bV_{\bY_{\textnormal{O}_2, r}}^{\top}$};
    \node[rotate=135] at (1.85-4.57, 3.0+4.2) {$\cdots$};

    \draw (2.80-4.58, 2.35+4.2) node[wide box] {};
    \node at (2.80-4.58, 2.35+4.2) {$\bV_{\bY_{\textnormal{H}_2\textnormal{O},r}}^{\top}$};

    \node at (0.75-5, 6.3+4.2) {$\bV_r^\top$};

    \draw (4.45-4.8, 7.2) node[long vec1] {};
    \draw (4.45-4.8, 5.7) node[hatched long box] {};
    \node at (4.45-4.80, 10.5) {$\widetilde{\bs}_k^*$};

    \node at (5.25-4.8, 4+4.2) {=};

    \draw (6.0-4.8, 4+4.2) node[short vec2] {};
    \draw (6.0-4.8, 3+4.2) node[short shaded vec] {};
    \node at (6.0-4.8, 6.3+4.2) {$\widehat{\bs}_k^*$};

    \node at (-3.3, 11.5) {\text{Projection of corrected high-dimensional state}};
    \node at (-3.3, 11.1) {\text{using a block-structured basis}};
\end{tikzpicture}
\caption{Interpretation of state-constraints-embedded state propagation, when using a block-structured POD basis. The left-hand side of the vertical dotted line represents the enforcement of species limiters, and the shaded area of $\widetilde{\bs}_k^*$ represents the corrected values. The right-hand side shows the projection of the partially corrected high-dimensional state, modifying only the entries of the reduced state $\widehat{\bs}_k^*$ corresponding to the mass fraction variables.}
\label{fig:block_structured_POD}
\end{figure}

\section{Numerical results} \label{sec:numerical_results}
This section presents the numerical results of our study on OpInf ROMs with state constraints for the char combustion problem.
\Cref{ss:FOM_data_set} presents the details of the MFiX-PIC FOM data from which the ROM is constructed.
In \Cref{ss:basis_selection}, we discuss the ROM dimension selection process.
\Cref{ss:regularization_hyperparameters_selection} presents the regularization hyperparameters selection process, where we propose a new selection criterion via a key performance indicator.
In \Cref{ss:various_methods}, we analyze the effect of enforcing state constraints to OpInf on stability, accuracy, and computational efficiency by comparing it with standard OpInf and other stability-enhancing OpInf approaches.
As discussed in Sections~\ref{sss:stabilizing_effect} and~\ref{sss:block_POD_basis}, the global basis provides stronger stabilizing effects due to the combination of state constraints and coupled ROM dynamics. Thus, for the results shown here, we use the global POD basis.

\subsection{FOM data details}    \label{ss:FOM_data_set}
The training data generated by MFiX-PIC consist of solution variables from Eq.~\eqref{eq:solution_variables} over $n_x = 22{,}400$ cells, with $N = n_v \times n_x = 224{,}000$ being the degrees of freedom of the high-fidelity system. The FOM simulation generates solutions over 38 seconds (from $t=2$ to $t=40$ seconds), using an adaptive time step ranging from $10^{-8}$s to $10^{-4}$s. The snapshots are saved every $\delta t=10^{-3}$s, providing $K_f=38{,}000$ snapshots. Of these, the initial $K=12{,}000$ snapshots (corresponding to $t \in [2,14]$s, $32\%$ of the data) are used for training, while the remaining $26{,}000$ snapshots (corresponding to $t \in [14,40]$s, $68\%$ of the data) are reserved for testing. Thus, the testing interval extends approximately $200\%$ beyond the length of the training interval. The high-dimensional data generation took approximately 56 days with an MPI-based solver on 45 CPUs, corresponding to 60,480 CPU hours (Intel Xeon W-3175X CPU @ 3.10 GHz). 
The construction of the OpInf ROM utilizes the training snapshot matrix $\bQ \in \real^{224{,}000 \times 12{,}000}$.~\Cref{tab:range_of_variable} shows the range of state variables within the training regime, which is essential for the data pre-processing step, as elaborated in~\Cref{ss:data_pre_processing}. The ranges and means of the state variables in the testing regime (not shown here) are similar to those in the training regime.

\subsection{Selection of the ROM dimension}    \label{ss:basis_selection}
Conventionally, POD considers an energy metric to determine the ROM dimension $r$. The cumulative energy $\cE_{\text{cum}}(r)$ is defined as the ratio of the sum of squared singular values up to rank $r$ to the total sum of squared singular values: $\cE_{\text{cum}}(r) =\sum_{\eta=1}^{r}\sigma_{\eta}^2 / \sum_{\eta=1}^{r_t}\sigma_{\eta}^{2}$, where the $\sigma_{\eta}$ ($\eta=1,2,\dots,r_t$) are the singular values of the scaled snapshot matrix $\bS$. Here, $r_t$ is the number of singular values to extract when we use randomized SVD in practice. This cumulative energy connects the low-rank dimension $r$ to the relative projection error of the scaled snapshot $\bS$ as
\begin{equation}
    \cE_{\text{proj}}(r) = \frac{\left\| \bS - \bV_r \bV_r^{\top} \bS \right\|_F^2}{\left\| \bS \right\|_F^2} = 1 - \cE_{\text{cum}}(r).
\label{eq:energy_criterion}
\end{equation}
For this data set, we found that $\cE_{\textnormal{proj}}(1) = 0.005$, meaning the cumulative energy is $0.995$ with $r=1$. Despite the first mode containing most of the system energy, choosing only one global mode for all variables is insufficient to capture the complex dynamics and sensitivities of the combustion model. \par

Instead of evaluating the single projection error of the scaled snapshot matrix $\bS$ that contains all the scaled variables, we can also calculate the projection error for each variable in its original scales, with respect to the low-dimensional rank-$r$ basis $\bV_r$. This involves reconstructing the scaled snapshots as $\widetilde{\bS} = \bV_r \bV_r^\top \bS$, followed by restoring the original scales of all $n_v = 10$ variables through the inversion of Eqs.~\eqref{eq:shifting}--\eqref{eq:scaling}. Denoting the reconstructed snapshot matrix of the $\ell$th state in its original scale as $\widetilde{\bQ}^{(\ell)} \in \real^{n_x \times K}$ (see \Cref{ss:data_pre_processing} for notation), the relative error is computed with respect to the true $\ell$th state snapshot matrix $\bQ^{(\ell)} \in \real^{n_x \times K}$.
\begin{figure}[ht]
    \begin{subfigure}[b]{0.50\linewidth}
        \centering
        \includegraphics[width=\linewidth]{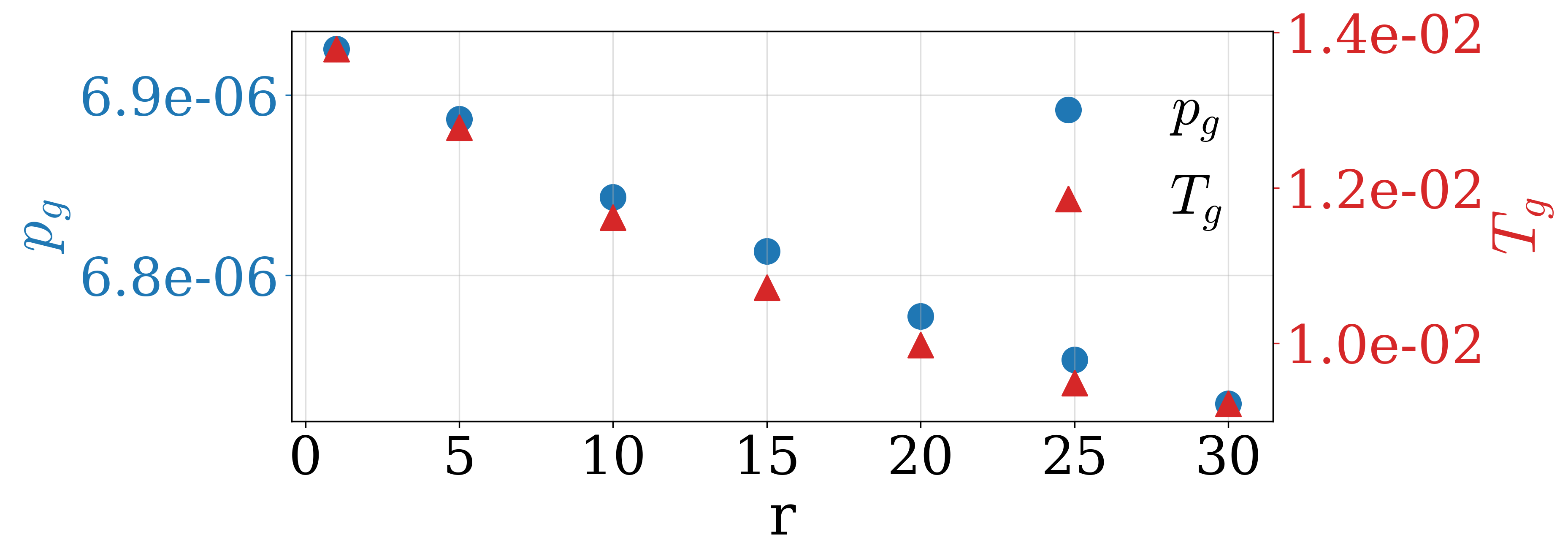}    
        \label{fig:projection_error_pT}
    \end{subfigure}
    \begin{subfigure}[b]{0.50\linewidth}
        \centering
        \includegraphics[width=\linewidth]{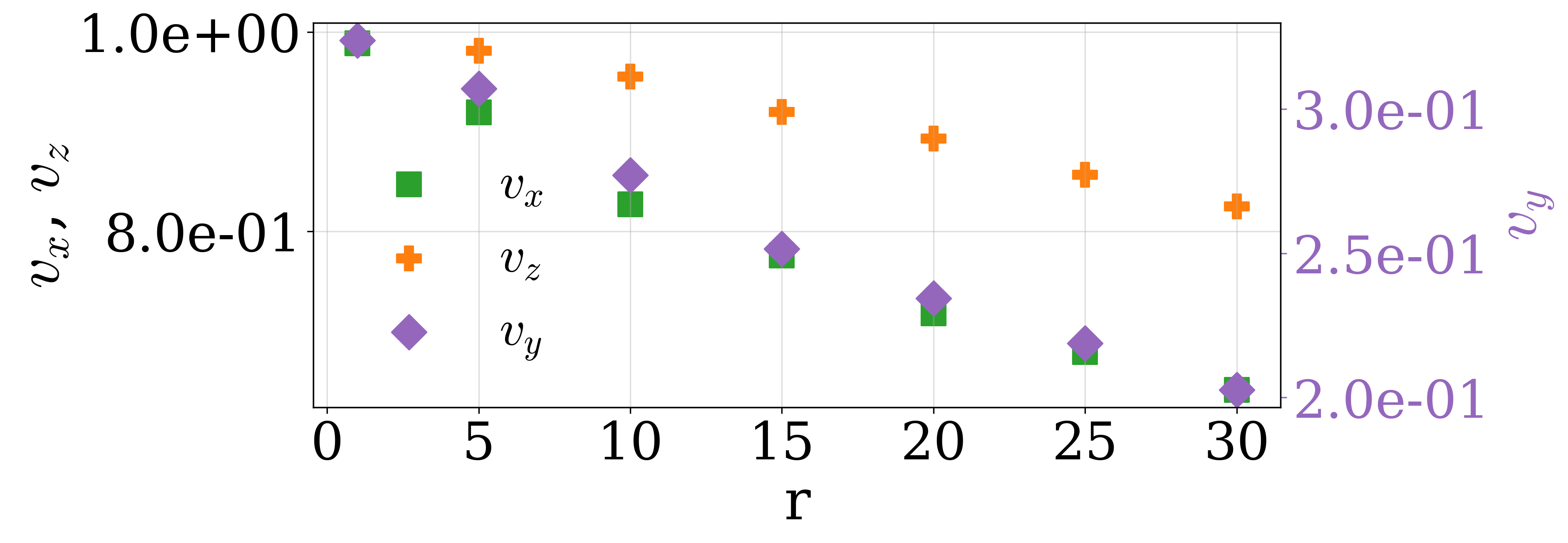}    
        \label{fig:projection_error_vels}
    \end{subfigure}
    \begin{subfigure}[b]{0.50\linewidth}
        \centering
        \includegraphics[width=\linewidth]{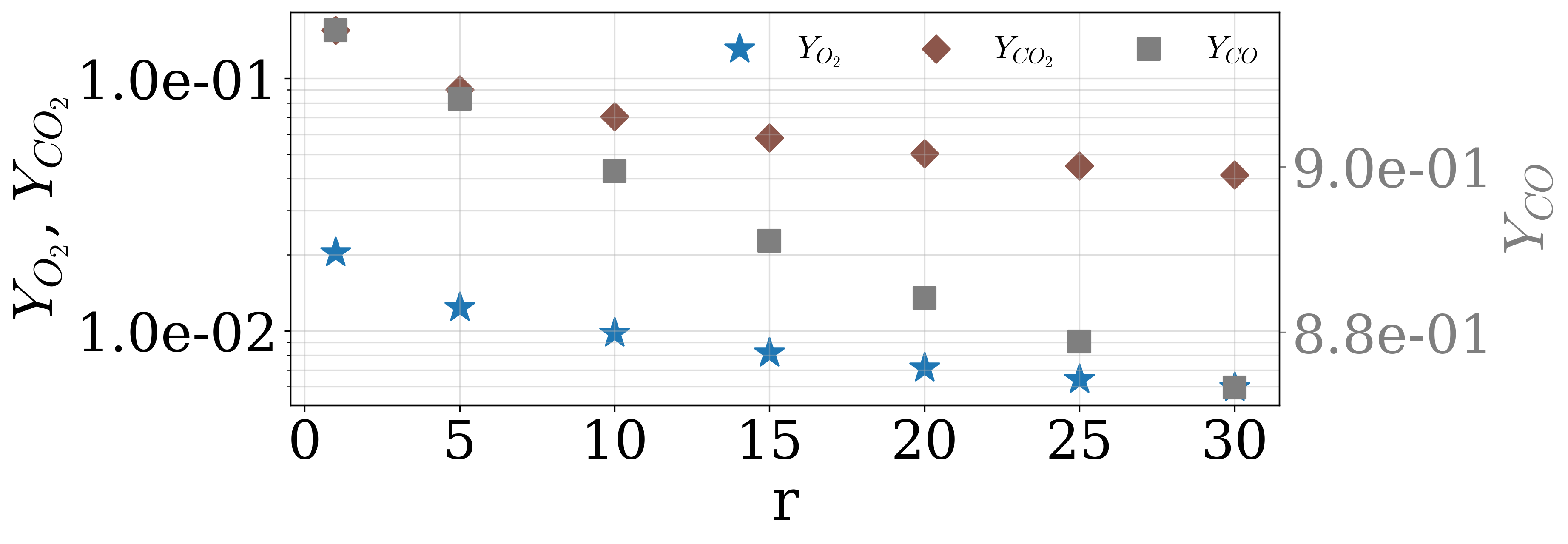}    
        \label{fig:projection_error_massfraction1}
    \end{subfigure}
    \hspace{+0.15em}
    \begin{subfigure}[b]{0.40\linewidth}
        \centering
        \includegraphics[width=\linewidth]{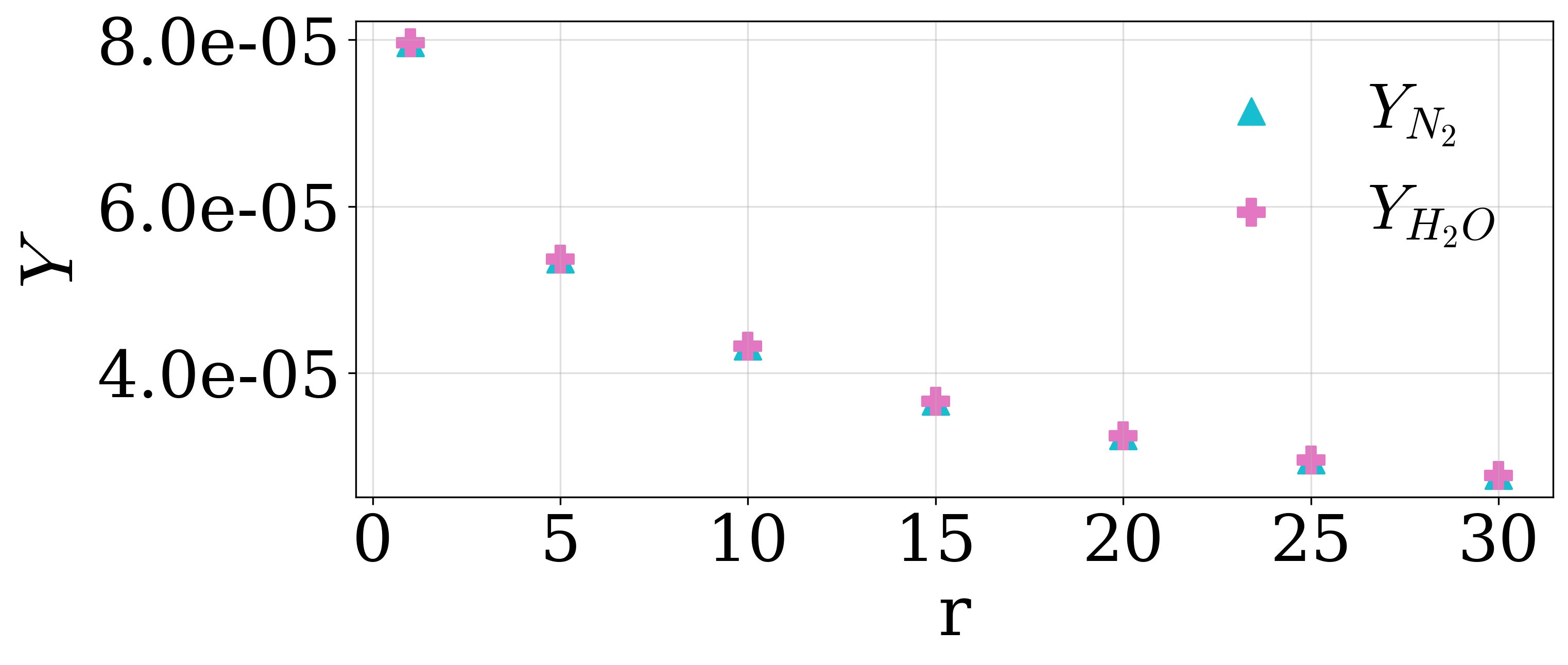}    
        \label{fig:projection_error_massfraction2}
    \end{subfigure}
    \vspace{-1.5em}
    \caption{Projection errors of each variable. The errors are computed with the variables in their original scales as $ \| \bQ^{(\ell)} - \widetilde{\bQ}^{(\ell)} \|_F^{2} / \| \bQ^{(\ell)} \|_F^{2}$.}
    \label{fig:projection_errors}
\end{figure}   

~\Cref{fig:projection_errors} shows the variable-specific projection errors in their original scales with respect to the low-rank dimension $r$. Pressure, $\textnormal{N}_2$, and $\textnormal{H}_2\textnormal{O}$ mass fractions exhibit notably low errors, in turn indicating high energy capture. Conversely, $\textnormal{CO}$ mass fraction and velocity components ($v_x$, $v_y$, $v_z$) display higher errors. This discrepancy is in part due to their relatively small scales, which skew the errors when considered in the denominator of the relative error metric. These observations highlight the strong dependence of projection errors on the specific variable under consideration. This variability suggests that the effectiveness of the rank-$r$ basis in capturing the system's dynamics is not uniform across all variables. \par

In conclusion, neither the projection error of the individual variables nor the energy associated with the entire scaled snapshot matrix $\bS$ provides a comprehensive representation of the complex and multiscale characteristics of this dataset.
After several tests, we select $r=30$ because it reasonably captures the high-frequency fluctuating dynamics of the FOM while providing a manageable model dimension. 
For this dataset, selecting more than $30$ modes is ineffective. This is because the regularization hyperparameter selection process, which is discussed next, substantially increases the computational time.

\subsection{Regularization hyperparameters selection based on key performance indicator}    \label{ss:regularization_hyperparameters_selection}
To select the hyperparameters $\lambda_1$ and $\lambda_2$ for regularizing the least-squares problem in Eq.~\eqref{eq:optimization_regularized}, we create a logarithmically-spaced uniform parameter grid, where $(\lambda_1, \lambda_2) \in [10^{-1}, 10^{6}] \times [10^{3}, 10^{10}]$, with each dimension discretized into 10 values. 
Using each set of inferred reduced operators with each regularization on the ($\lambda_1, \lambda_2$) grid, the ROM propagates the initial reduced state $\widehat{\bs}_0$ within the training regime through numerical time integration of Eq.~\eqref{eq:polynomial_ROM_eq}, obtaining the ROM solution in the reduced space $\widehat{\bS}^{\textnormal{ROM}} = \left[ \widehat{\bs}_{0}^{\textnormal{ROM}}, \cdots, \widehat{\bs}_{K-1}^{\textnormal{ROM}} \right] \in \real^{r \times K}$.
Conventionally, one selects the best $\lambda_1^*$ and $\lambda_2^*$ values that minimize the relative state error between the FOM reduced states Eq.~\eqref{eq:low_dim_snapshots} and the ROM solution,
\begin{align}
    e^{\textnormal{ROM}} = \frac{\| \widehat{\bS}^{\textnormal{FOM}} - \widehat{\bS}^{\textnormal{ROM}} \|_F}{\|\widehat{\bS}^{\textnormal{FOM}}\|_F} \times 100~(\%).
\label{eq:state_error}
\end{align} 
While widely used, this approach is inadequate in our case because it measures accuracy only in \emph{the scaled reduced state space}, where contributions of different variables are scaled and mixed. As a result, a model may achieve a lower relative state error yet fail to capture essential nonlinear dynamics in the physical variables. \par
\definecolor{black}{rgb}{0, 0, 0} 
\definecolor{red}{rgb}{1, 0, 0}   
\definecolor{blue}{rgb}{0, 0, 1}  
\begin{figure}[ht]
    \centering
    \begin{minipage}{1.0\textwidth}
    \centering
    \begin{tabular}{lll}
        \textcolor{black}{\rule{1.5em}{1.0pt}} FOM &
        \textcolor{blue}{\tikz\draw[dash dot, line width=0.4mm] (0,0) -- (0.7,0);} ROM based on Eq.~\eqref{eq:state_error} &
        \textcolor{red}{\tikz\draw[dotted, line width=0.4mm] (0,0) -- (0.7,0);} ROM based on KPI criteria
    \end{tabular}
    \vspace{0.3em}
    \end{minipage}
    \begin{subfigure}[b]{0.45\linewidth}
        \centering
        \includegraphics[width=\linewidth]{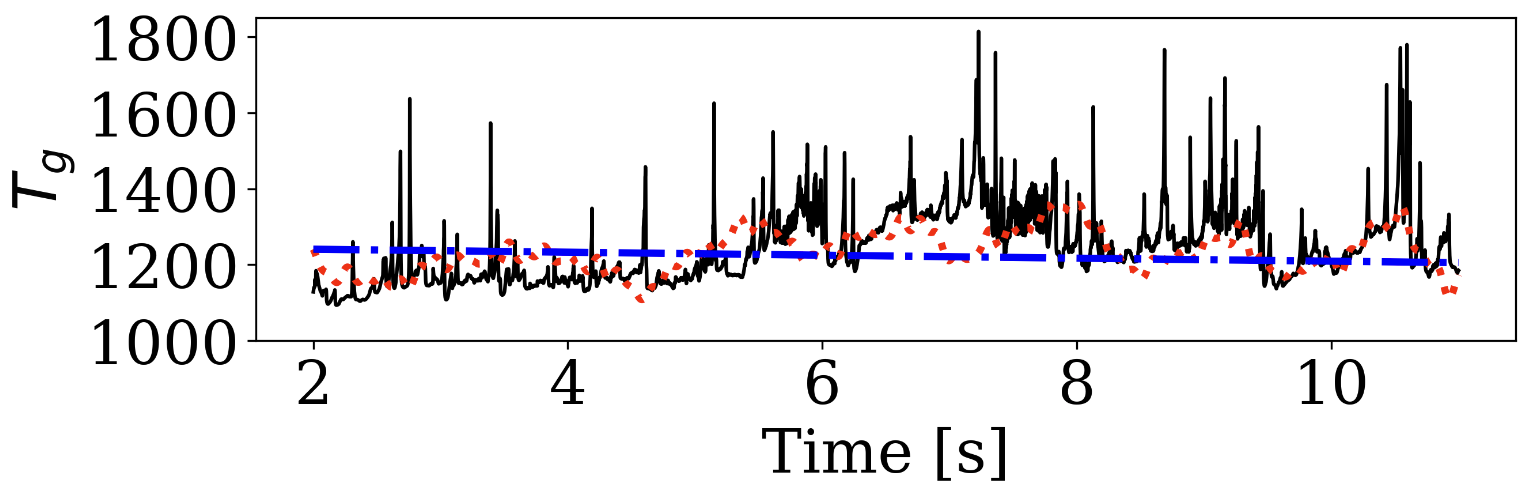}    
        \label{fig:temperature_comparison}
    \end{subfigure}
    \begin{subfigure}[b]{0.45\linewidth}
        \centering
        \includegraphics[width=\linewidth]{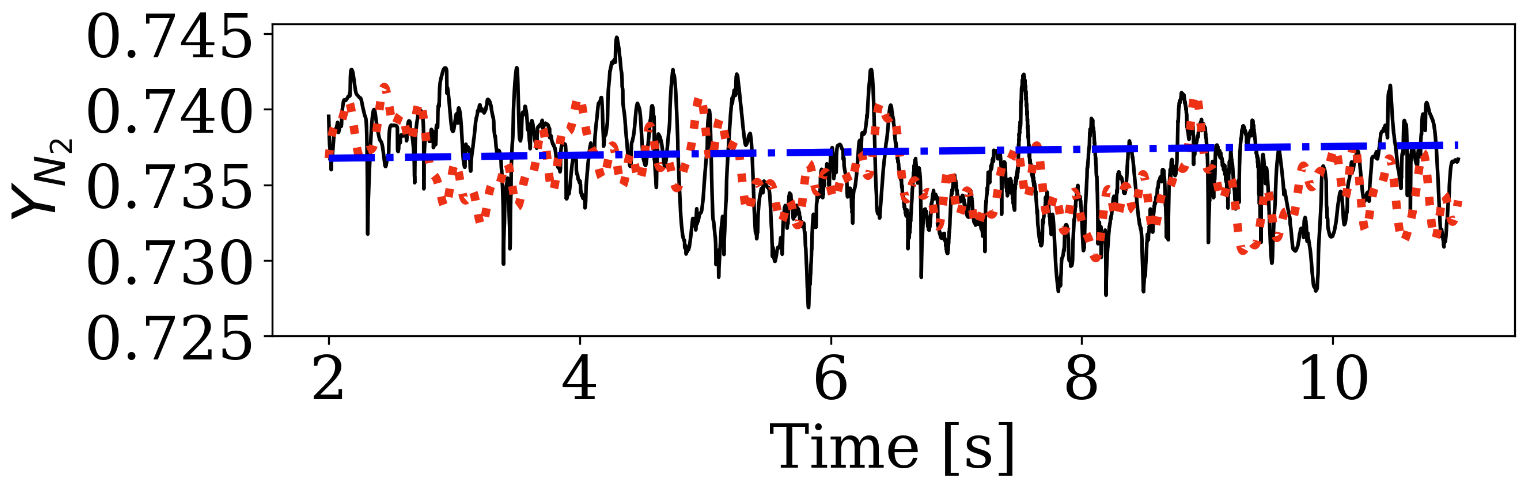}    
        \label{fig:N2_mass_fraction_comparison}
    \end{subfigure}
    \vspace{-1.5em}
    \caption{Comparison of two ROM predictions of temperature and $\textnormal{N}_2$ mass fraction over 9{,}000 snapshots spanning from 2 to 11 seconds, averaged across the boiler outlet surface. 
    }
    \label{fig:grid_search_comparison}
\end{figure}
We propose to use the key performance indicator of char combustion as an alternative way of selecting hyperparameters for regularization. During each ROM prediction in the ($\lambda_1$, $\lambda_2$) grid search, we compute a crucial derived quantity of char combustion as the KPI: the thermal energy collected at the boiler outlet $A_O$ (see~\Cref{fig:computational_domain}). This quantity serves as a ROM selection metric throughout the grid search.
The thermal energy $\xi(t) [\textnormal{J}]$ is formulated as
\begin{equation}
    \xi(t) = \int_{t_0}^{t}
    \underbrace{\left[\frac{1}{|A_O|}\int_{A_O}c_{p,g}(\bx,\tau)\cdot \dot{m}(\bx,\tau)\cdot \Delta T(\bx,\tau)\mathrm{d}\bx \right]
    }_{\dot{\xi}(\tau)}
    \mathrm{d}\tau.
\label{eq:thermal_energy_eq}
\end{equation}
Here, the specific heat capacity of the gas mixture, $c_{p,g}$, is computed as $\sum_{i} {c_{p,i} \cdot X_i/ M_{\textnormal{mix}}}$ where $M_\textnormal{mix} = \sum_{i} X_i \cdot M_i$ represents molar mass of the gas mixture. The molar fraction of each species is $X_i$, and its molar mass is $M_i$, i.e., $M_1=31.9988 [\textnormal{g}/\textnormal{mol}]$ $(\textnormal{O}_2)$, $M_2=28.0134 [\textnormal{g}/\textnormal{mol}]$ $(\textnormal{N}_2)$, $M_3 = 28.0104 [\textnormal{g}/\textnormal{mol}]$ $(\textnormal{CO})$, $M_4 = 44.0098 [\textnormal{g}/\textnormal{mol}]$ $(\textnormal{CO}_2)$, and $M_5=18.0153 [\textnormal{g}/\textnormal{mol}]$ $(\textnormal{H}_2\textnormal{O})$. The mass flow rate, $\dot{m}$ $[\textnormal{kg}/\textnormal{s}]$, through a surface is computed by summing the product of the gas phase volume fraction, density, velocity component normal to the boiler outlet surface $A_O$, and the area: $\int \varepsilon_g \cdot \rho_g \cdot v_y \cdot \mathrm{d} A_O$, which in turn is computed in the discretized setting as $\sum_{ijk} \varepsilon_{g, ijk} \cdot \rho_{g,ijk} \cdot v_{y, ijk} \cdot |A_{O,ijk}|$, where $ijk$ are the spatial indices of the discretization of $A_O$.
Here, $\dot{\xi}(\tau)$ represents the thermal energy rate $[\textnormal{J}/\textnormal{s}]$, averaged across the boiler outlet $A_O$.
To compute Eq.~\eqref{eq:thermal_energy_eq}, all state variables except for $v_x$ and $v_z$ are required. However, the computation only needs the reconstruction of the high-dimensional variables in a specific region, $A_O$, making it computationally feasible. \par

We generate two ROMs based on different regularization methods: one is chosen via the KPI-based approach in Eq.~\eqref{eq:thermal_energy_eq}, where thermal energy guides hyperparameter selection, and the ROM that is trained by minimizing the relative reduced-state error of Eq.~\eqref{eq:state_error}.
\Cref{fig:grid_search_comparison} compares temperature and $\textnormal{N}_2$ mass fraction predictions for those two ROMs. It demonstrates that our proposed KPI-based approach provides a more physically accurate ROM for this problem, as it captures key nonlinear combustion dynamics in both temperature and $\textnormal{N}_2$ mass fraction better than the other ROM, which is not expressive despite minimizing~Eq.~\eqref{eq:state_error}. In particular, the conventional relative state error metric of Eq.~\eqref{eq:state_error} assigns the KPI-based ROM a slightly lower score of $e^{\textnormal{ROM}} = 6.64\%$ compared to $e^{\textnormal{ROM}} = 6.35\%$ for the other that minimizes Eq.~\eqref{eq:state_error}. This highlights that minimizing the reduced-state error does not ensure physical fidelity for our complex, multiscale char combustion case. In contrast, the proposed KPI-based approach aggregates contributions from multiple variables in the \emph{full physical space}, directly reflecting the underlying combustion physics, and yields a ROM that better preserves critical combustion behavior.
\par
For this dataset, the proposed approach is less expensive than the conventional method of selecting regularization hyperparameters based on the state error minimization metric. This is primarily because the norm computation for $\|\widehat{\bS}-\widehat{\bS}^{\textnormal{ROM}}\|_F / \|\widehat{\bS}\|_F$ in the conventional method is costly as the number of training snapshots $K$ is high. When computing the error of the predicted KPI, we compare its full time series of the thermal energy $\xi(t)$, not the thermal energy accumulated at the final training time step, $\xi(t_{K-1})$. This avoids cases where the final energy would be similar, but the production curves thereof would have differed significantly. 
We also exclude regularizations that produce negative values of $\dot{\xi(t)}$.
For each combination of regularizations, the OpInf problem is solved, the resulting ROM is integrated in time, and $\xi(t)$ is computed from the reconstructed high-dimensional states over a partial domain---combustion outlet $A_O$---using the unconstrained ROM trajectory. Note that state constraints are not applied during this step, thereby avoiding the substantial offline cost of repeatedly enforcing them in the full high-dimensional space throughout the grid search.
The hyperparameters, $(\lambda_1, \lambda_2) = (129.1550, 5994.8425)$, are selected via this KPI-based grid search.
These hyperparameters minimize the error between $\xi^{\textnormal{FOM}}(t)$ and $\xi^{\textnormal{ROM}}(t)$, resulting in a relative error of $10.57 \%$. 
%

\subsection{Comparison with other stability-enhancing Operator Inference reduced-order model approaches}    \label{ss:various_methods}
Since the proposed OpInf with state constraints also enhances the stability of the ROM, we compare it with two other advanced stability-enhancing approaches of OpInf: eigenvalue reassignment and constrained optimization. Additional results for OpInf using log-transformed species mass fractions for positivity preservation are included in~\ref{app:log_transformation}.

\subsubsection{Eigenvalue reassignment} \label{sss:eigenvalue_reassignment}
The OpInf framework finds the reduced operators by minimizing the cost function of Eq.~\eqref{eq:optimization_regularized}. As a result, it does not guarantee that the inferred linear operator $\widehat{\bA} \in \real^{r \times r}$ has only stable eigenvalues. Eigenvalue reassignment generates a stable $\widehat{\bA}$ by repositioning its unstable eigenvalues to be stable (see, e.g.,~\cite[Sec.~3.4.1]{SAWANT2023115836}). First, assuming $\widehat{\bA}$ is diagonalizable, we identify the eigenvalues of $\widehat{\bA}$ by computing its eigendecomposition as
\begin{equation}
    \widehat{\bA} = \mathbf{\Phi} \mathbf{\Sigma}_{\widehat{\bA}} \mathbf{\Phi}^{-1}
\label{eq:eigval_reassignment}
\end{equation}
where the columns of $\mathbf{\Phi}$ are eigenvectors of $\widehat{\bA}$, and $\mathbf{\Sigma}_{\widehat{\bA}}$ is a diagonal matrix containing the eigenvalues of $\widehat{\bA}$.
Assume that the leading $p$ eigenvalues of $\widehat{\bA}$ have non-negative real parts. Let these eigenvalues be $\sigma_1, \dots, \sigma_p$ and let the remaining eigenvalues be $\sigma_{p+1}, \dots, \sigma_{r}$. Denote $\Re(\sigma)$ and $\Im(\sigma)$ as real and imaginary parts of the eigenvalue $\sigma$. We replace $\Re(\sigma_1), \dots, \Re(\sigma_p)$ with a small tolerance $-\epsilon$, where $\epsilon > 0$.
The linear reduced operator $\widehat{\bA}$ is then replaced by a new stable matrix $\mathbf{\Phi} \widetilde{\mathbf{\Sigma}} \mathbf{\Phi}^{-1}$, where $\widetilde{\mathbf{\Sigma}} = \textnormal{diag}(-\epsilon + \Im(\sigma_1), \dots, -\epsilon + \Im(\sigma_p), \sigma_{p+1}, \dots, \sigma_{r})$. As a result, all the eigenvalues in $\widetilde{\mathbf{\Sigma}}$ have strictly negative real parts.

\subsubsection{Operator Inference with constrained optimization}    \label{sss:constrained_optimization}
The authors in~\cite{SAWANT2023115836} considered constrained optimization within a structure-preserving OpInf framework to identify a symmetric negative definite linear matrix $\widehat{\bA}$. 
They demonstrated the local stabilization of the OpInf ROM by constrained optimization on Burgers' equation and the reaction-diffusion equation.
Yet, its performance in strongly nonlinear multiphysics problems, with conditions not close to the linear regime, is not certain.
For the purpose of the comparison in this paper, we solve the constrained optimization for the OpInf with char combustion data.
For numerical purposes, we implement this with a model constraint $\widehat{\bA}+\epsilon \bI \preceq \textbf{0}$ in Eq.~\eqref{eq:optimization_regularized}, thereby finding a stable matrix $\widehat{\bA}$. Here, $\bI \in \real^{r \times r}$ is an identity matrix, $\widehat{\bA}+\epsilon \bI \preceq \textbf{0}$ means $\widehat{\bA}+\epsilon \bI$ is negative semidefinite, and the tolerance $\epsilon > 0$ ensures that $\widehat{\bA}$ is negative definite.
 We use the \texttt{cvxpy 1.6} python package~\cite{diamond2016cvxpy, agrawal2018rewriting} for negative semidefinite programming, without a symmetricity constraint and with a default tolerance of $\epsilon = 10^{-8}$.

\subsubsection{Error metric}    \label{sss:error_metric}
We calculate pointwise errors for all variables in their original scales as an error metric. Since pressure and temperature have large magnitudes, we use a standard relative error for these variables. The notation follows Eq.~\eqref{eq:l_th_state}, except that the prediction is made for the $K_f=38{,}000$ snapshots for each $\ell$th state variable on the $n_x=22{,}400$ cells, such that $\bQ^{(\ell), .} \in \real^{22{,}400 \times 38{,}000}$. The relative pointwise error is computed as
$$
\Upsilon^{\textnormal{rel}}_{j,k} = \frac{\left | \bQ^{(\ell),\textnormal{FOM}}_{j,k} - \bQ^{(\ell),\textnormal{ROM}}_{j,k} \right |}{\left | \bQ^{(\ell), \textnormal{FOM}}_{j,k} \right |},
$$
where $j$ and $k$ denote spatial and temporal indices.
However, this error metric can often skew a small error if true values in the denominator are small. Thus, for other variables (velocity components and species mass fractions) that have relatively smaller scales and even include values close to zero, we compute a normalized absolute error
$$
\Upsilon^{\textnormal{nabs}}_{j,k} = \frac{\left | \bQ^{(\ell), \textnormal{FOM}}_{j,k} - \bQ^{(\ell), \textnormal{ROM}}_{j,k} \right |}{\textnormal{max} \left( \left | \bQ^{(\ell), \textnormal{FOM}} \right | \right)}.
$$
Here, the denominator denotes the maximum of absolute values of $\bQ^{(\ell), \textnormal{FOM}}$ throughout the entire spatial and temporal domain.
The resulting error matrix $\Upsilon \in \real^{n_x \times K_f}$ contains pointwise errors for all $K_f$ snapshots throughout the training and testing regime, over the entire spatial domain. To analyze the temporal evolution of the error across the spatial domain, we calculate the spatially averaged pointwise error where each element $\overline{\Upsilon}_k = \frac{1}{n_x} \sum_{j=1}^{n_x} \Upsilon_{j,k}$ represents the spatially averaged pointwise error at the time step $k$.

\subsubsection{Results} \label{sss:results}
\begin{figure}[htbp]
    \begin{minipage}{0.4\textwidth}
        \centering
        \begin{tabular}{ll}
            \textcolor{C01}{\tikz\draw[line width=0.4mm] (0,0) -- (0.7,0);} Standard OpInf &
            \textcolor{C06}{\tikz\draw[dashed, line width=0.4mm] (0,0) -- (0.7,0);} OpInf with eigenvalue reassignment \\
            \textcolor{C02}{\tikz\draw[dotted, line width=0.5mm] (0,0) -- (0.7,0);} OpInf with constrained optimization &
            \textcolor{blue}{\tikz\draw[dash dot, line width=0.4mm] (0,0) -- (0.7,0);} OpInf with species limiters
        \end{tabular}
        \vspace{0.3em}
    \end{minipage}

    \begin{subfigure}[b]{0.95\linewidth}
        \centering
        \includegraphics[width=\linewidth]{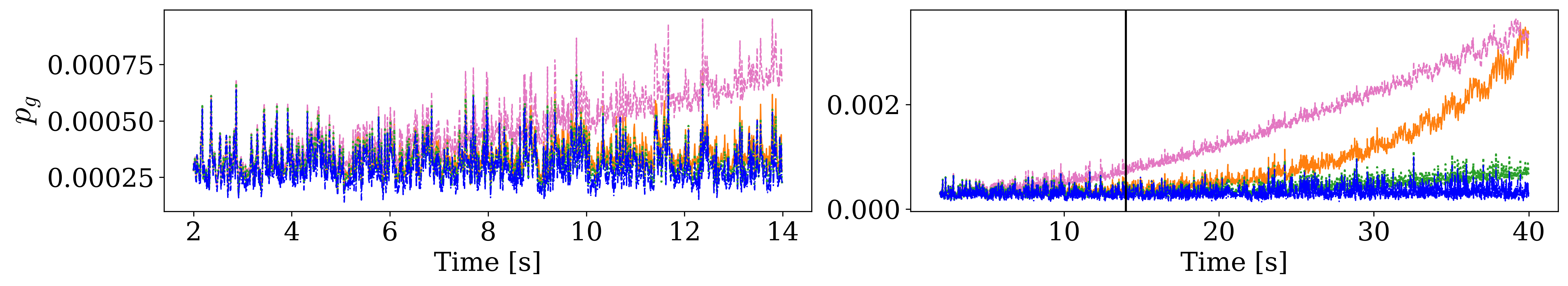}
        \label{fig:error_p}
    \end{subfigure}
    
    \begin{subfigure}[b]{0.95\linewidth}
        \centering
        \includegraphics[width=\linewidth]{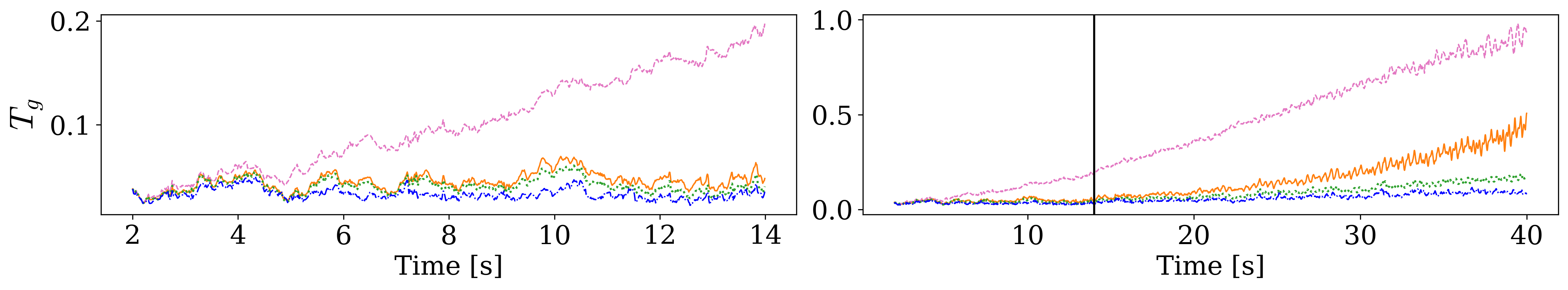}
        \label{fig:error_T}
    \end{subfigure}
        
    \begin{subfigure}[b]{0.95\linewidth}
        \centering
        \includegraphics[width=\linewidth]{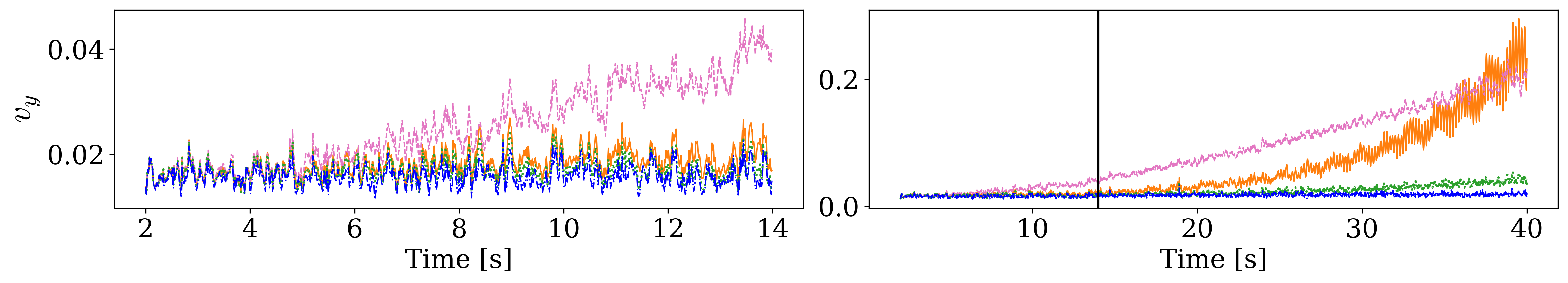}
        \label{fig:error_vy}
    \end{subfigure}
    
    \begin{subfigure}[b]{0.95\linewidth}
        \centering
        \includegraphics[width=\linewidth]{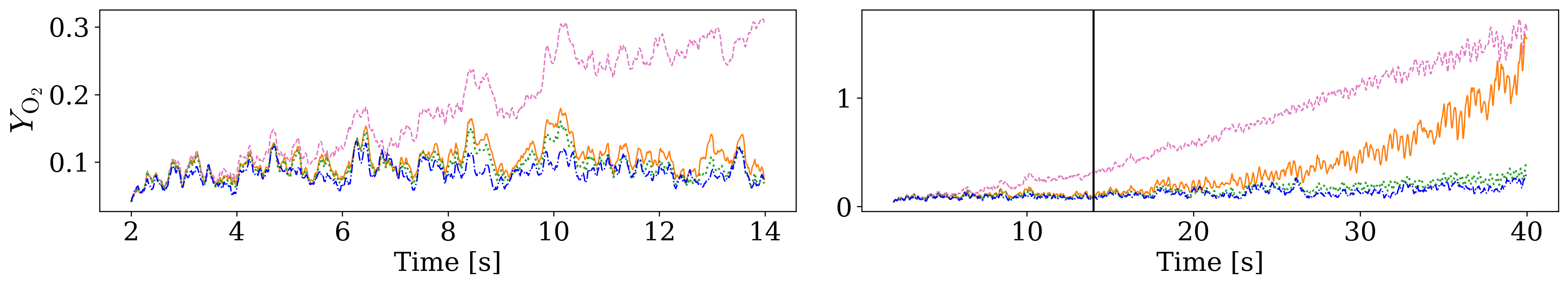}
        \label{fig:error_O2}
    \end{subfigure}
    
    \begin{subfigure}[b]{0.95\linewidth}
        \centering
        \includegraphics[width=\linewidth]{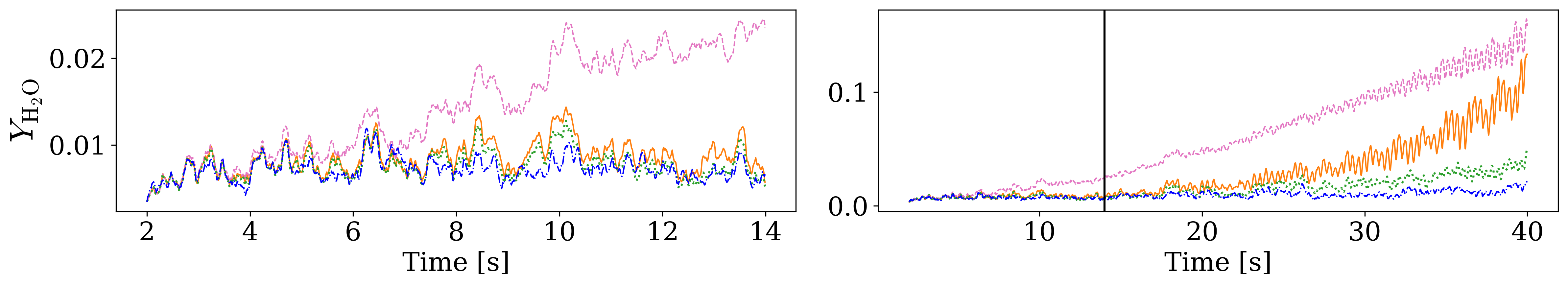}
        \label{fig:error_H2O}
    \end{subfigure}

    \caption{Comparison of spatially averaged pointwise error $\overline{\Upsilon}_k$ of all methodologies. Figures in the left column show error propagation over the training regime, while those in the right column show the full time horizon. The black vertical lines mark the end of the training regime, which comprises $12{,}000$ snapshots in $t \in [2,14]$s, followed by the testing regime ($26{,}000$ snapshots in $t \in [14,40]$s). Variables not shown here are presented in~\ref{app:error_all}.}
    \label{fig:all_error_comparison}
\end{figure}

\paragraph{Stability and error comparison}    \label{par:error}
~\Cref{fig:all_error_comparison} compares errors introduced in~\Cref{sss:error_metric} for four models: (i) the proposed OpInf with species limiters, (ii) standard (unconstrained) OpInf, (iii) OpInf with eigenvalue reassignment, (iv) OpInf with constrained optimization. All ROMs propagate low-dimensional states using the RK45 time-integration scheme.
For a fair comparison and to isolate the effect of each OpInf variant, we apply the same regularization hyperparameters, $\lambda_1$ and $\lambda_2$, across all OpInf ROMs. These regularizers are selected based on the KPI criterion for the standard OpInf. This ensures that any observed performance differences arise solely from the formulation of each OpInf variant, rather than from differences in regularization.
Except for the proposed OpInf with species limiters, the other ROMs implemented propagate states without state constraints, see also~\Cref{fig:species_limiter_step}, where this is illustrated with dashed arrows.
\par
First, the error evolution of all OpInf models is similar in the training regime, except for OpInf with eigenvalue reassignment. This reflects the contribution of appropriate regularization to accuracy, achieved through the KPI-based selection of hyperparameters during the model learning phase.
As the predictions go beyond the training regime, we observe a divergence of errors for all variables with the standard OpInf. In contrast, the OpInf with state constraints on the species mass fractions maintains stable and consistently lower errors. This demonstrates that enforcing state constraints enhances accuracy by providing a stabilizing effect.
The OpInf with eigenvalue reassignment produces larger errors than the standard OpInf within the training regime, eventually resulting in divergent errors in the testing regime.
This highlights that making a highly nonlinear system locally stable simply by reassigning unstable eigenvalues of its linear operator does not guarantee global ROM stability and accuracy. In addition, the divergent predictions arise because the quadratic term in the model Eq.~\eqref{eq:polynomial_ROM_eq} dominates the system dynamics in the testing regime. 
In constrast, the OpInf with constrained optimization learns the optimal stable $\widehat{\bA} \prec \textbf{0}$ in the model learning process. Its error behavior in the training regime is similar to the OpInf with species limiters. 
In the testing regime, its errors become slightly larger than those of the OpInf with species limiters. Note that although evaluating errors after $t=40$ seconds is not possible because we do not have FOM data, we can still check if the ROM predictions are stable and physically consistent. We found that predictions by OpInf with constrained optimization begin to diverge at $t=50$ seconds, whereas OpInf with species limiters is stable yet becomes less expressive.
Similar to the OpInf with eigenvalue reassignment, the divergent ROM predictions by OpInf with constrained optimization in the time horizon far beyond the testing regime are due to the predominant influence of the quadratic term in the ROM.
We also investigated cubic OpInf ROMs with KPI-based regularization. However, their offline training---especially the grid search for optimal regularization hyperparameters---was substantially more expensive than that of quadratic ROMs due to the increased number of operations (see~\Cref{ss:OpInf_regularization}). In addition, the cubic ROMs yielded less expressive solutions than the quadratic ROMs.
\par
A critical observation is that the state constraints applied to the species mass fractions provide a global stabilizing effect across all variables, including pressure, temperature, and velocity components that are not subject to constraints. This added benefit is due to using a global POD basis. As depicted in~\Cref{fig:state_correction_illustration} and detailed in~\Cref{sss:stabilizing_effect}, at each time step $k$, the entire reduced state vector $\widehat{\bs}_k$ is \textit{modified} to $\widehat{\bs}_k^*$, even though the constraints are applied only to species mass fractions. The modified reduced state $\widehat{\bs}_k^*$ evolves to $\widehat{\bs}_{k+1}$. The $\widehat{\bs}_{k+1}$ is then lifted to the high-dimensional scaled state $\widetilde{\bs}_{k+1}$, which contains all variables. This process ensures that the ROM always propagates entirely modified reduced states, eventually influencing the trajectories of all variables.
\par

\paragraph{Conservation of physical constraints}    \label{par:unphysical_predictions}
Next, we check if the OpInf ROMs make unphysical predictions.~\Cref{tab:invalid_species_numbers} presents the ratio of the unphysical species mass fractions predicted with each OpInf ROM over the entire spatial and temporal regime. Predictions on species mass fractions are considered unphysical if they violate the conditions specified in Eqs.~\eqref{eq:species_limiter_O2}--\eqref{eq:species_limiter_H2O}. The standard OpInf, OpInf with eigenvalue reassignment, and OpInf with constrained optimization generate unphysical predictions for all species, whereas the OpInf with species limiters~Eqs.~\eqref{eq:species_limiter_O2}--\eqref{eq:species_limiter_H2O} remains physically consistent for all species across the entire spatio-temporal domain. If we further relax the upper bounds in Eqs.~\eqref{eq:species_limiter_O2},~\eqref{eq:species_limiter_N2}, and~\eqref{eq:species_limiter_H2O} to one (which would be known a priori without FOM or the boundary condition data of this specific model), the ROM produces unphysical results for $\textnormal{O}_2$, $\textnormal{N}_2$, and $\textnormal{H}_2\textnormal{O}$.
This demonstrates that incorporating any available physical information into the design of state constraints improves the physical consistency of the ROM predictions.
We note that, while the KPI strategy improves model accuracy during training, it alone does not ensure physically valid predictions: the physically consistent behavior of species-limited OpInf can be directly attributed to the enforcement of state constraints.
We also observe unphysical predictions by OpInf with species limiters enforced every $\alpha$th time step, see the last two rows of~\Cref{tab:invalid_species_numbers}. We provide a detailed discussion together with the computational time analysis presented later.
\begin{table}[htbp]
    \centering
    \caption{The percentage (\%) of unphysical species mass fractions in $n_x = 22{,}400$ spatial grid cells over $K_f=38{,}000$ snapshots. The last three cases apply the species limiters~Eqs.~\eqref{eq:species_limiter_O2}--\eqref{eq:species_limiter_H2O} every $\alpha$th time step.}
    \vspace{-0.5em}
    \begin{tabular}{l|r r r r r}
        & $\textnormal{O}_2$ & $\textnormal{N}_2$ & $\textnormal{CO}$ & $\textnormal{CO}_2$ & $\textnormal{H}_2\textnormal{O}$ \\
        \hline
        Standard OpInf & 26.74 & 26.57 & 30.79 & 18.95 & 26.57 \\
        OpInf with eigenvalue reassignment & 45.28 & 40.20 & 46.78 & 25.64 & 40.20 \\
        OpInf with constrained optimization & 8.30 & 18.76 & 18.66 & 6.98 & 18.76 \\
        OpInf with species limiters of a generic bound [$0, 1$] & 2.64 & 20.06 & 0 & 0 & 20.06 \\
        OpInf with species limiters ($\alpha=1$) & 0 & 0 & 0 & 0 & 0 \\
        OpInf with species limiters ($\alpha=10$) & 0.64 & 0.15 & 6.35 & 0.86 & 0.15 \\
        OpInf with species limiters ($\alpha=100$) & 3.11 & 2.65 & 15.23 & 3.30 & 0.65
    \end{tabular}
    \label{tab:invalid_species_numbers}
\end{table}
\par
\begin{figure}[ht]
    \begin{minipage}{0.4\textwidth}
        \centering
        \begin{tabular}{ll}
            \textcolor{C01}{\tikz\draw[line width=0.4mm] (0,0) -- (0.7,0);} Standard OpInf &
            \textcolor{C06}{\tikz\draw[dashed, line width=0.4mm] (0,0) -- (0.7,0);} OpInf with eigenvalue reassignment \\
            \textcolor{C02}{\tikz\draw[dotted, line width=0.5mm] (0,0) -- (0.7,0);} OpInf with constrained optimization &
            \textcolor{blue}{\tikz\draw[dash dot, line width=0.4mm] (0,0) -- (0.7,0);} OpInf with species limiters
        \end{tabular}
        \vspace{0.3em}
    \end{minipage}

    \centering
    \includegraphics[width=0.6\linewidth]{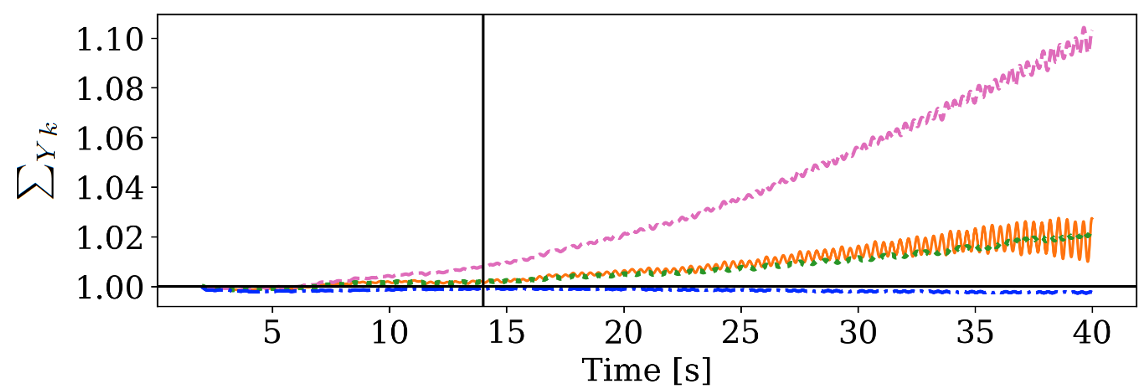}
    \vspace{-0.4em}
    \caption{The sum of mass fractions over time. The sum is computed by summing five mass fractions and then averaging over the spatial discretization dimension: ${\sum_{Y}}_k = \frac{1}{n_x} \sum_{j=1}^{n_x} {\bY_{\textnormal{O}_2}}_{j,k} + {\bY_{\textnormal{N}_2}}_{j,k} + {\bY_{\textnormal{CO}}}_{j,k} + {\bY_{\textnormal{CO}_2}}_{j,k} + {\bY_{\textnormal{H}_2\textnormal{O}}}_{j,k}$, where $j$ and $k$ represent spatial and temporal indices.}
    \label{fig:sum_mass_fractions}
\end{figure}

Additionally, we check whether each ROM conserves the sum of mass fractions, $Y_{\textnormal{O}_2} + Y_{\textnormal{N}_2} + Y_{\textnormal{CO}} + Y_{\textnormal{CO}_2} + Y_{\textnormal{H}_2\textnormal{O}}$, ensuring that it remains equal to 1. \Cref{fig:sum_mass_fractions} compares the spatially averaged sum of mass fractions over time for each ROM. Among the four methods, only OpInf with species limiters consistently maintains the sum of five mass fractions close to 1 across the entire spatio-temporal domain. Note that the proposed method does not explicitly enforce this constraint; however, the species limiters defined in Eq.~\eqref{eq:species_limiter_O2}--\eqref{eq:species_limiter_H2O} help preserve this property.
\par

\begin{figure}[htbp]
\begin{minipage}{0.3\textwidth}
    \centering
    \begin{tabular}{ll}
        \textcolor{black}{\rule{2em}{1.0pt}} FOM &
        \textcolor{C01}{\tikz\draw[dashed, line width=0.4mm] (0,0) -- (0.7,0);} Standard OpInf \\
        \textcolor{C06}{\tikz\draw[dashed, line width=0.4mm] (0,0) -- (0.7,0);} OpInf with eigenvalue reassignment &
        \textcolor{C02}{\tikz\draw[dotted, line width=0.5mm] (0,0) -- (0.7,0);} OpInf with constrained optimization \\
        \textcolor{blue}{\tikz\draw[dash dot, line width=0.4mm] (0,0) -- (0.7,0);} OpInf with species limiters
    \end{tabular}
    \vspace{0.3em}
\end{minipage}
    
\begin{subfigure}[b]{0.49\textwidth}
    \centering
    \includegraphics[width=\textwidth]{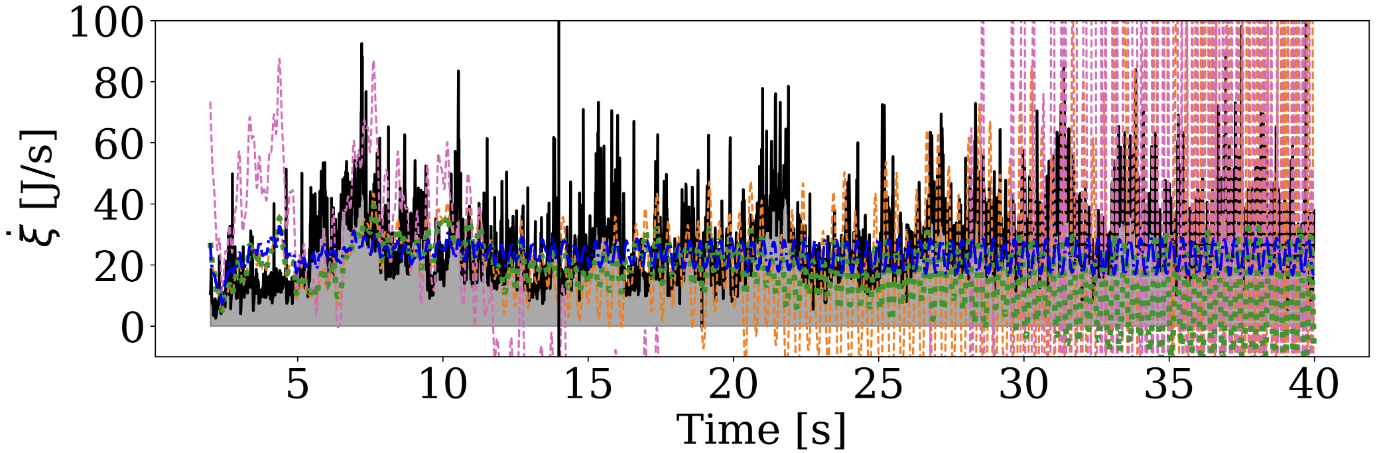}
    \caption{Thermal energy rate}
    \label{fig:TE_rate}
\end{subfigure}
\hfill
\begin{subfigure}[b]{0.49\textwidth}
    \centering
    \includegraphics[width=\textwidth]{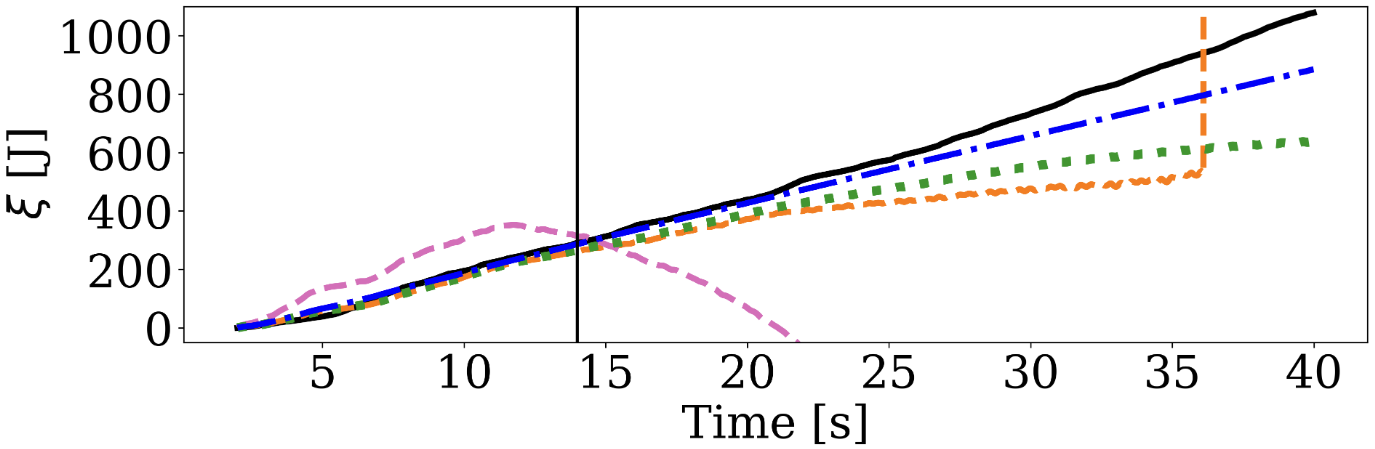}
    \caption{Thermal energy}
    \label{fig:TE_propagation}
\end{subfigure}

\caption{Comparison of thermal energy rate and thermal energy. The shaded area under the reference $\dot{\xi}$, representing its time-integrated quantity, corresponds to the thermal energy $\xi$.}
\label{fig:thermal_E_comparison}
\end{figure}

\paragraph{Thermal energy prediction} \label{par:KPI_performance}
~\Cref{fig:thermal_E_comparison} presents another comparison of the four methods regarding the prediction capabilities for the thermal energy, our KPI. 
As shown in~\Cref{fig:TE_rate}, the standard OpInf, OpInf with eigenvalue reassignment, and OpInf with constrained optimization yield unstable predictions for the thermal energy rate, $\dot{\xi}$, computed using Eq.~\eqref{eq:thermal_energy_eq}. Notably, these methods even predict negative values for $\dot{\xi}$ in the testing regime.
This is unphysical because $\dot{\xi}$ must inherently remain strictly positive, given that the specific heat capacity $c_{p,g}$, mass flow rate $\dot{m}$, and the temperature gradient $\Delta T$ are all positive over the boiler outlet $A_O$). 
In particular, the OpInf with eigenvalue-reassignment predicts negative $\dot{\xi}$ even in the training regime, despite using the same regularizers as the other methods. This stems from the fact that the unstable eigenvalues of $\widehat{\bA}$ are reassigned after the model is trained.
The instability and inaccuracy $\dot{\xi}(t)$ in the three methods lead to correspondingly unstable and inaccurate thermal energy predictions in~\Cref{fig:TE_propagation}.
On the other hand, OpInf with species limiters makes stable and accurate predictions throughout training and testing, resulting in the closest predictions to the time series of the FOM thermal energy.
\par
\paragraph{Comparison of computational time}   \label{par:computational_time_result}
~\Cref{tab:computational_time} compares the four methods' computational cost. We run all four ROMs using Python 3.12.2 on an MPI-based cluster equipped with 56 Intel Xeon W-3175X CPUs @ 3.10 GHz. The offline and online computational costs are measured in wall clock time. 
The online speedup factor is computed by dividing the FOM simulation cost (60,480 CPU hours) by the ROM online cost, with all ROM costs converted to CPU hours.
Specifically, we measure the ROM online CPU time using the \texttt{time.process\_time()} function in \texttt{time} module of Python. The online phase for all methods includes the temporal evolution of the reduced states by the ROMs in the entire time domain, followed by the projection of the results into the high-dimensional space.
The offline phase for all methods comprises data pre-processing, randomized SVD, and solving the optimization problem with regularization to learn the ROMs. The randomized SVD, which takes approximately 2.42 minutes in wall time, is the most expensive step in the offline phase. For the OpInf with eigenvalue reassignment, the offline phase also includes reallocating unstable eigenvalues of the learned linear operator $\widehat{\bA} \in \real^{r \times r}$. This process is executed instantaneously, as the dimension of $\widehat{\bA}$ is only $r \ll N$.
\begin{table}[ht]
    \centering
    \caption{Comparison of computational wall time for the offline stage (model learning in the training regime) and the online stage (time integration in the entire time regime) across all methods. The last three cases apply the species limiters every $\alpha$th time step to investigate possible avenues for online speedup. To account for variations in ROM simulation time, we average the costs over 10 runs for each method.}
    \vspace{-0.5em}
    \begin{tabular}{l|r r r}
         & Offline [mins] & Online [mins] & Online speedup factor \\
        \hline
        Standard OpInf & 2.68 & 0.01 & $4.41\times10^9$ \\
        OpInf with eigenvalue reassignment & 2.69 & 0.01 & $4.41\times10^9$ \\
        OpInf with constrained optimization & 64.36 & 0.01 & $4.41\times10^9$ \\
        OpInf with species limiters ($\alpha=1$) & 2.68 & 20.83 & $3.17\times10^3$ \\
        OpInf with species limiters ($\alpha=10$) & 2.68 & 5.14 & $1.26\times10^4$ \\
        OpInf with species limiters ($\alpha=100$) & 2.68 & 1.64 & $3.95\times10^4$
    \end{tabular}
    \label{tab:computational_time}
\end{table}
The offline stage of the OpInf with constrained optimization includes solving a regularized optimization problem with a model constraint $\widehat{\bA}+\epsilon \bI \preceq \textbf{0}$, which can be expensive depending on the amount of training data and the ROM dimension. In our case, this process takes more than 60 minutes on average over 10 runs in wall time.
Lastly, one should note that all four models are trained on the same amount of high-fidelity data with the same projection matrix. If not otherwise available, simulating that data may also factor into the offline cost. \par

\begin{figure}[ht]
    \centering
    \begin{minipage}{1.0\textwidth}
    \centering
    \begin{tabular}{llll}
        \tikz{\draw[C00, thick, dashed] (0,0) -- (0.8,0);} $K=5{,}000$ &
        \tikz{\draw[C06, thick, dash dot] (0,0) -- (0.8,0);} $K=9{,}000$ &
        \textcolor{C08}{\rule{1.5em}{1.0pt}} $K=12{,}000$ &
        \tikz{\draw[C09, thick, dotted] (0,0) -- (0.8,0);} $K=15{,}000$
    \end{tabular}
    \vspace{0.3em}
    \end{minipage}

    \begin{center}
    \begin{minipage}{0.48\textwidth}
        \centering
        Standard OpInf
    \end{minipage}%
    \hspace{2em} 
    \begin{minipage}{0.4\textwidth}
        \centering
        OpInf with species limiters
    \end{minipage}
    \end{center}
    
    \begin{subfigure}[b]{1.0\linewidth}
        \centering
        \includegraphics[width=\linewidth]{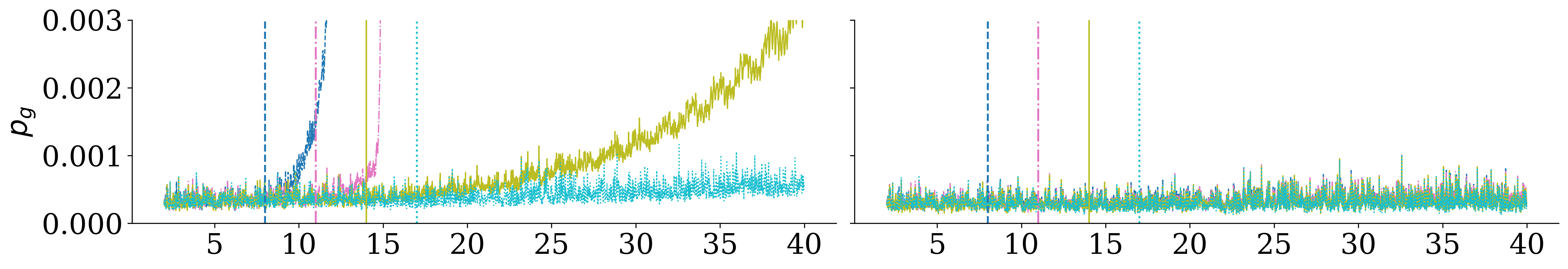}    
        \label{fig:different_training_pressure}
    \end{subfigure}
    \vspace{-1em}
    \begin{subfigure}[b]{1.0\linewidth}
        \centering
        \includegraphics[width=\linewidth]{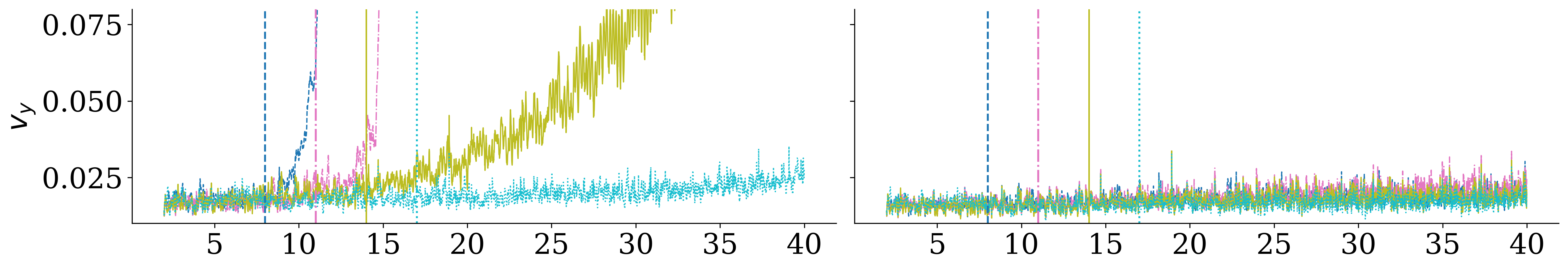}
        \label{fig:different_training_vy}
    \end{subfigure}
    \vspace{-1em}
    \begin{subfigure}[b]{1.0\linewidth}
        \centering
        \includegraphics[width=\linewidth]{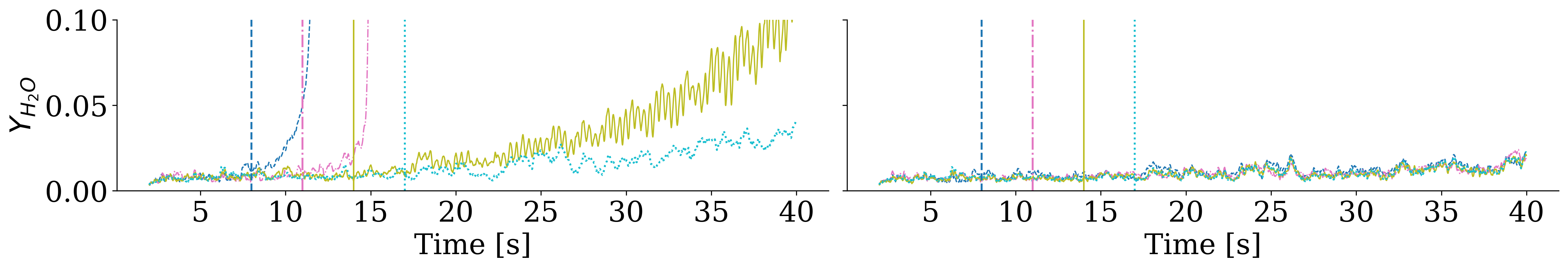}
        \label{fig:different_training_H2O}
    \end{subfigure}
    \caption{Comparison across four training horizons: $2$–$7\text{s}$ ($K=5{,}000$ snapshots), $2$–$11\text{s}$ ($K=9{,}000$), $2$–$14\text{s}$ ($K=12{,}000$), and $2$–$17\text{s}$ ($K=15{,}000$). The vertical lines indicate the end of each training regime, with the line color and style matching the corresponding error curve.}
    \label{fig:different_training_comparison}
\end{figure}

\paragraph{Sensitivity of OpInf with species limiters to the amount of training data}
\label{par:different_training_time}
We investigate the influence of training data length on the performance of the OpInf ROM.~\Cref{fig:different_training_comparison} shows the comparison of error propagation for standard OpInf ROMs and species-limited OpInf ROMs trained with varying numbers of snapshots, with all models selected based on the KPI-based regularization with parameters given in~\Cref{ss:regularization_hyperparameters_selection}.
The standard OpInf shows unstable error growth when trained with $K=5{,}000$, $9{,}000$, and $12{,}000$ snapshots, whereas extending the training set to $15{,}000$ snapshots improves stability and extends accuracy further into the testing horizon. In contrast, the species-limited OpInf ROMs yield consistently stable error behavior across all training lengths. The error curves for these cases nearly overlap, indicating that the performance of the species-limited OpInf ROMs is largely insensitive to the amount of training data.
\par

\paragraph{Summary of ROM performance}  \label{par:summary_of_ROMs}
The performance of the four OpInf ROMs can be summarized as follows. First, standard OpInf, OpInf with eigenvalue reassignment, and OpInf with constrained optimization are the fastest methods online. However, they encounter stability and accuracy issues when making long-term predictions, often resulting in physically inconsistent results and inaccurate predictions on our KPI.
OpInf with constrained optimization involves considerable offline computational costs, making it more expensive to train compared to other approaches.
Additionally, this method does not always produce expressive ROMs when evaluated with training data across different temporal regimes. For example, using a
larger amount of training data (e.g., $K = 15{,}000$ snapshots) results in less expressive predictions with
fewer oscillations compared to the FOM data. Conversely, unlike the species-limited ROM cases shown in~\Cref{fig:different_training_comparison}, training with fewer snapshots (e.g., $K=5{,}000$ and $K=9{,}000$ snapshots) leads to instability, even with stabilization efforts.
Lastly, OpInf that enforces species limiters every time step incurs the highest online cost, leading to the smallest online speedup among all methods. This is because it requires reconstructing high-dimensional states at each time step, enforcing species mass fraction constraints (as outlined in Eqs.~\eqref{eq:species_limiter_O2}--\eqref{eq:species_limiter_H2O}), and projecting the modified states back to the low-dimensional space, see~\Cref{fig:species_limiter_step}. 
Nevertheless, it achieves a significant speedup of $3{,}170$ times compared to the FOM computation.
While enforcing species limiters every $\alpha$th time step (here, $\alpha=10$ and $100$) improves online speedup, as shown in~\Cref{tab:invalid_species_numbers}, this strategy does not ensure physically valid mass fraction predictions at intermediate time steps when the constraints are not applied.
As demonstrated in the numerics above, only OpInf with species limiters at every time step ensures physically consistent predictions for species mass fractions in this application. It also provides the best accuracy and stability among all methods tested. \par

\section{Conclusions}        \label{sec:conclusion}
We proposed a new approach to enhance accuracy and stability of nonintrusive OpInf ROMs, while ensuring physical consistency, as demonstrated in a char combustion application. We incorporated state constraints into the prediction step of the OpInf ROMs, specifically for the species mass fraction variables, enabling physically consistent predictions across the entire spatio-temporal domain. Our approach combines efficient ROM state evolution with an online correction that partially reintroduces FOM information, forming a hybrid offline-online framework.
We further introduced a strategy for selecting regularization hyperparameters in OpInf ROM training, guided by key performance indicators (KPIs) that can be defined for a wide range of physical and engineering applications.
The proposed OpInf ROM with state constraints demonstrated superior performance in terms of stability, accuracy, and physical consistency, compared to the standard OpInf and other stability-enhancing approaches, which failed to fully preserve species mass fraction bounds and showed error growth, albeit at different rates.
We further showed that combining a global POD basis with state constraints provides the strongest stabilizing effect and the highest prediction accuracy, outperforming the block-POD basis. This approach also stabilizes other variables, such as pressure, temperature, and velocity components, which are not directly constrained.
Although the state-constrained OpInf ROM incurs additional online computational cost, it still achieved a significant speedup of approximately 3,170 times relative to the FOM simulation. \par

The proposed approach offers a promising avenue for developing fast, stable, and physically consistent data-driven ROMs for large-scale nonlinear systems. Future work could explore formulating the OpInf framework in an integral or a discrete-time form to address challenges in estimating time derivatives of POD coefficients for highly noisy, oscillatory, and coarsely sampled data. Also, investigating ways to reduce the online computational cost of the current expensive state constraints step, which relies on FOM information in the high-dimensional space, would be very valuable. We also attempted to impose limiting conditions directly on the reduced states after projecting the species mass fraction variable onto its own latent space via a block-structured basis. However, the physical bounds are encoded through coupled interactions between POD modes and coefficients, making it difficult to enforce such constraints in the low-dimensional space. Finally, the very recent work~\cite{BAUMGART2024113199} proposes a method to ensure the sum of mass fractions is equal to 1 in high-fidelity simulations of reacting flows. It would be interesting to impose such a constraint with our proposed state constraints in the ROM.

\section*{CRediT authorship contribution statement}
\textbf{Hyeonghun Kim}: Conceptualization, Data curation, Formal analysis, Methodology, Software, Validation, Visualization, Writing – original draft, Writing – review \& editing.~\textbf{Boris Kramer}: Conceptualization, Formal analysis, Funding acquisition, Methodology, Project administration, Supervision, Writing – review \& editing.

\section*{Declaration of competing interest}
The authors declare that they have no known competing financial interests or personal relationships that could have appeared to influence the work reported in this paper.

\section*{Data availability}
Data will be made available on request.

\section*{Acknowledgments}
This research was in part financially supported by the Korea Institute for Advancement of Technology (KIAT) through the International Cooperative R\&D program (No. P0019804, Digital twin-based intelligent unmanned facility inspection solutions).

\newpage
\appendix

\section{Nomenclature}
\label{app:nomenclature}
\begin{framed}
\begin{multicols}{2}
\begin{description}[leftmargin=!, labelwidth=0.8cm]
\item[$\rho_g$] gas density [$\textnormal{kg}/\textnormal{m}^3$]
\item[$\bv_{g}$] gas velocity vector in 3D space [$\textnormal{m/s}$]
\item[$\bv_{p}$] solid particle velocity vector in 3D space [$\textnormal{m/s}$]
\item[$N_{g}$] number of chemical species in gas mixture
\item[$R_{g,i}$] rate of formation per unit volume of $i$th gas species
\item[$R_{p,i}$] rate of formation for the $p$th MFiX-PIC parcel of $i$th chemical species
\item[$S_{g}$] gas source term [$\textnormal{kg}/(\textnormal{m}^3\cdot\textnormal{s})$]
\item[$S_{p}$] solid source term [$\textnormal{kg}/(\textnormal{m}^3\cdot\textnormal{s})$]
\item[$Y_{i}$] mass fraction of $i$th gas species
\item[$Y_{p,i}$] mass fraction of $i$th gas species for the $p$th parcel
\item[$X_{i}$] molar fraction of $i$th gas species
\item[$\bj_{g}$] diffusive flux vector
\item[$p_{g}$] gas pressure [$\textnormal{Pa}$]
\item[$\bg$] gravitational acceleration vector [$\textnormal{m}/\textnormal{s}^2$]
\item[$\mu_g$] dynamic viscosity [$\textnormal{kg}/(\textnormal{m}\cdot \textnormal{s})$]
\item[$\bS_{g}$] general gas phase momentum source term vector [$\textnormal{kg}/(\textnormal{m}^2\cdot\textnormal{s}^2)$]
\item[$\boldsymbol{\tau}_g$] gas stress tensor [$\textnormal{kg}/(\textnormal{m}\cdot \textnormal{s}^2$)]
\item[$\boldsymbol{\tau}_p$] interparticle stress tensor [$\textnormal{kg}/(\textnormal{m}\cdot \textnormal{s}^2$)]
\item[$c_{p,g}$] gas phase mixture specific heat capacity at constant pressure [$\textnormal{J/(kg} \cdot \textnormal{K)}$]
\item[$c_{p,p}$] solid parcel specific heat capacity at constant pressure [$\textnormal{J/(kg} \cdot \textnormal{K)}$] 
\item[$c_{p,i}$] $i$th gas specific heat capacity [$\textnormal{J/(kg} \cdot \textnormal{K)}$]
\item[$T_{g}$] gas phase temperature [$\textnormal{K}$]
\item[$T_{p}$] solid phase temperature [$\textnormal{K}$]
\item[$\kappa$] thermal conductivity [$\textnormal{W/(m} \cdot \textnormal{K)}$]
\item[$h_{p,i}$] $i$th species specific enthalpy [$\textnormal{J}/\textnormal{kg}$]
\item[$m_{\textnormal{ci}}$] mass of unreacted char [$\textnormal{kg}$]
\item[$d_{p,i}$] diameter of a char particle [$\textnormal{m}$]
\item[$m_p$] particle mass [$\textnormal{kg}$]
\item[$p_{\textnormal{oxy}}$] partial pressure of oxygen in gas mixture [$\textnormal{Pa}$]
\item[$M$] molecular weight [$\textnormal{g}/\textnormal{mol}$]
\item[$R_{\textnormal{diff}}$] diffusion rate
\item[$R_{\textnormal{chem}}$] kinetic rate
\item[$Sh$] Sherwood number
\item[$D_{o}$] binary diffusion coefficient of oxygen-nitrogen mixture [$\textnormal{m}^2/\textnormal{s}$]
\item[$R_u$] universal gas constant [$\textnormal{J}/(\textnormal{kmol}\cdot \textnormal{K})$]
\item[$T_m$] mean temperature between gas and solid [$\textnormal{K}$]
\item[$A_{\textnormal{pre}}$] pre-exponential factor
\item[$E_0$] activation energy
\item[$Re_{g}$] Reynolds number
\item[$\overline{U}$] slip velocity between fluid and particles
\item[$Sc$] Schmidt number
\item[$\varepsilon_g$] gas volume fraction
\item[$W_p$] statistical weight of particles
\end{description}
\end{multicols}
\end{framed}

\section{Error propagation of all variables}
\label{app:error_all}
To complement the results presented in~\Cref{fig:all_error_comparison}, we include here the error plots of the remaining five variables. These results verify the same overall trends of error evolution discussed in~\Cref{sss:results}.
\begin{figure}[htbp]
    \begin{minipage}{0.4\textwidth}
    \centering
    \begin{tabular}{ll}
        \textcolor{C01}{\tikz\draw[line width=0.4mm] (0,0) -- (0.7,0);} Standard OpInf &
        \textcolor{C06}{\tikz\draw[dashed, line width=0.4mm] (0,0) -- (0.7,0);} OpInf with eigenvalue reassignment \\
        \textcolor{C02}{\tikz\draw[dotted, line width=0.5mm] (0,0) -- (0.7,0);} OpInf with constrained optimization &
        \textcolor{blue}{\tikz\draw[dash dot, line width=0.4mm] (0,0) -- (0.7,0);} OpInf with species limiters
    \end{tabular}
    \vspace{0.3em}
    \end{minipage}

    \begin{subfigure}[b]{0.95\linewidth}
        \centering
        \includegraphics[width=\linewidth]{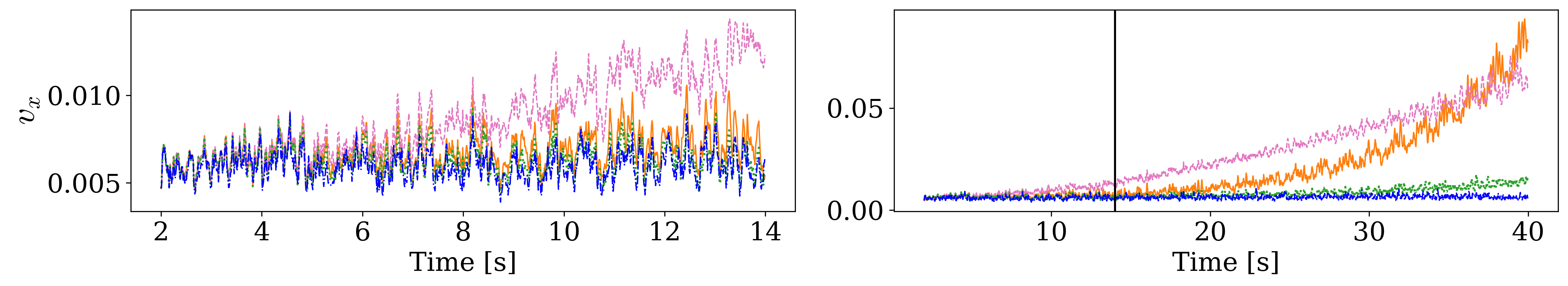}
        \label{fig:error_vx}
    \end{subfigure}
    
    \begin{subfigure}[b]{0.95\linewidth}
        \centering
        \includegraphics[width=\linewidth]{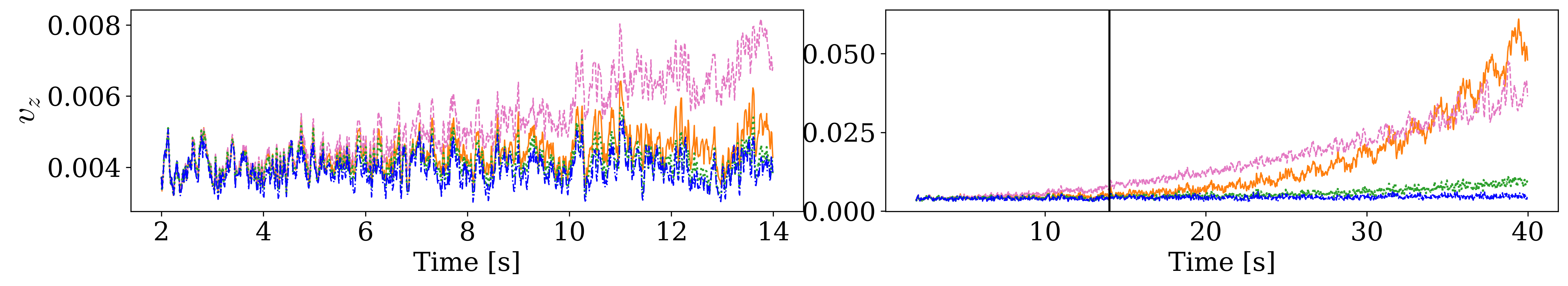}
        \label{fig:error_vz}
    \end{subfigure}
    
    \begin{subfigure}[b]{0.95\linewidth}
        \centering
        \includegraphics[width=\linewidth]{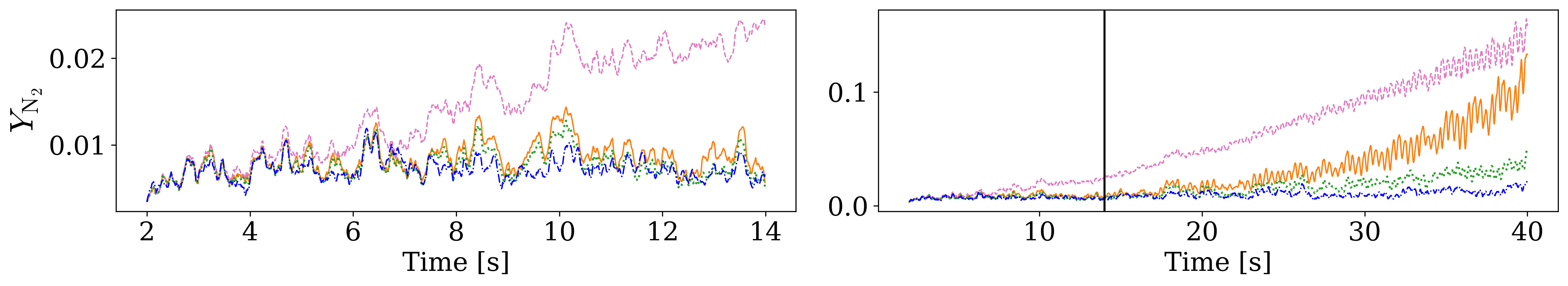}
        \label{fig:error_N2}
    \end{subfigure}

    \begin{subfigure}[b]{0.95\linewidth}
        \centering
        \includegraphics[width=\linewidth]{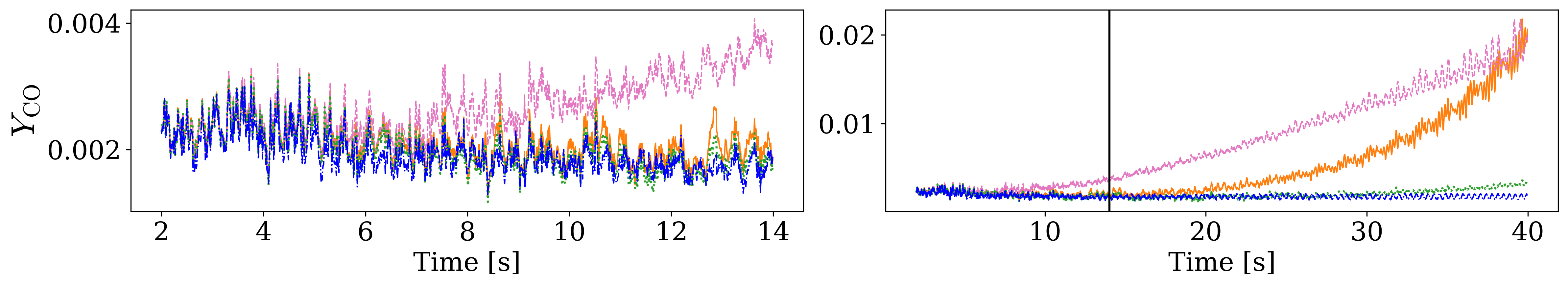}
        \label{fig:error_CO}
    \end{subfigure}
    
    \begin{subfigure}[b]{0.95\linewidth}
        \centering
        \includegraphics[width=\linewidth]{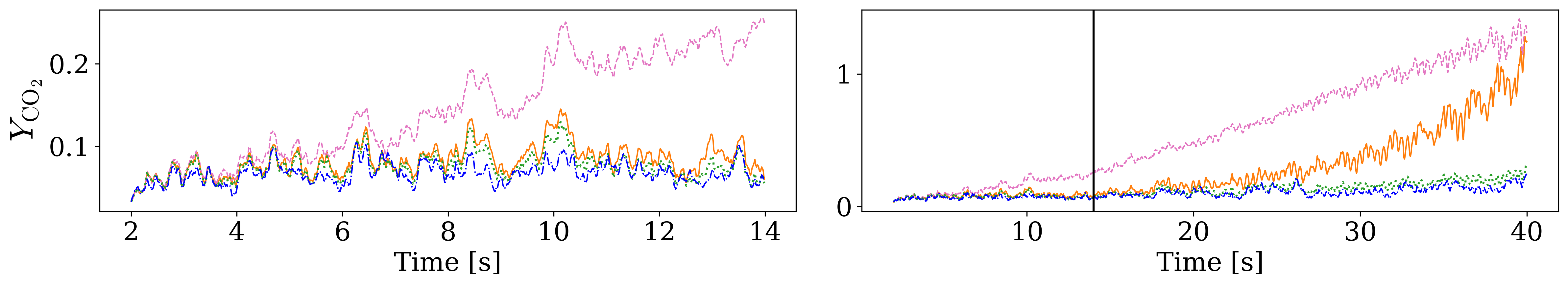}
        \label{fig:error_CO2}
    \end{subfigure}
    \vspace{-1em}
    \caption{Figures in the left column show error propagation over the training regime, while those in the right column show the error propagation over the full time horizon.}
    \label{fig:error_appendix}
\end{figure}

\section{Operator Inference using log-transformed species mass fractions for positivity preservation}
\label{app:log_transformation}
\begin{figure}[htbp]    
\begin{subfigure}[b]{0.49\textwidth}
    \centering
    \includegraphics[width=\textwidth]{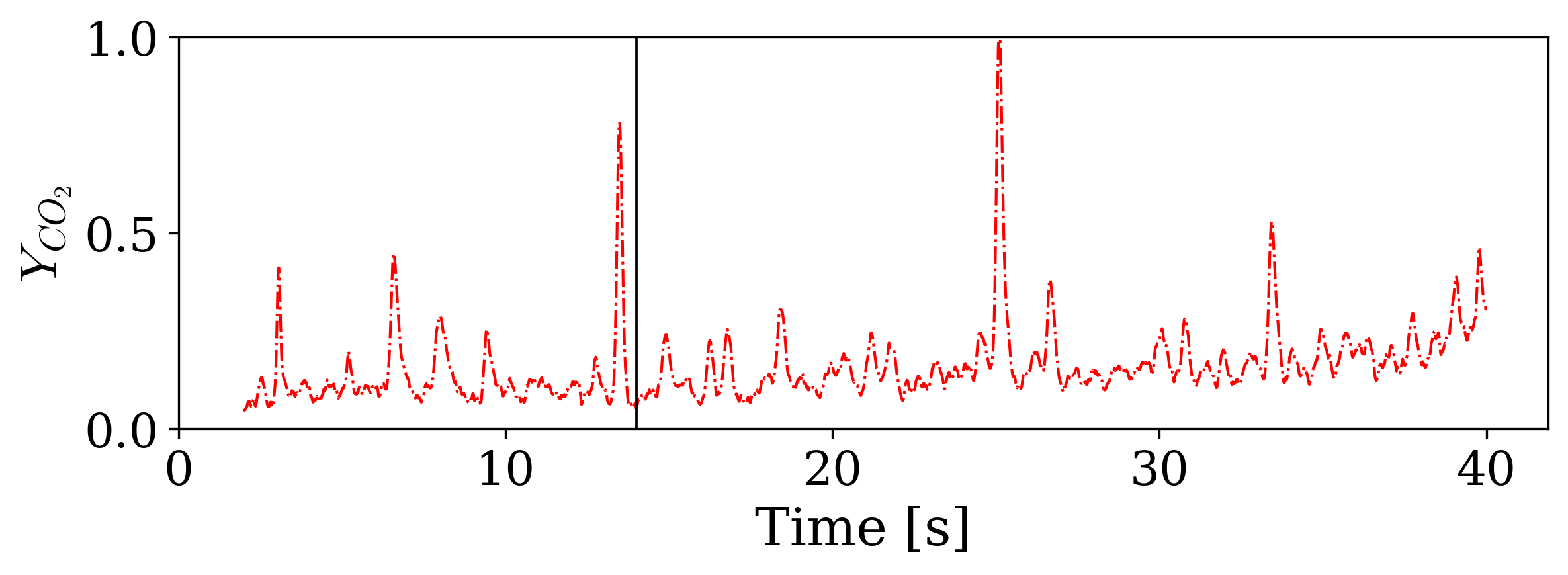}
    \vspace{-1.7em}
    \label{fig:error_log_CO2}
\end{subfigure}
\hfill
\begin{subfigure}[b]{0.49\textwidth}
    \centering
    \includegraphics[width=\textwidth]{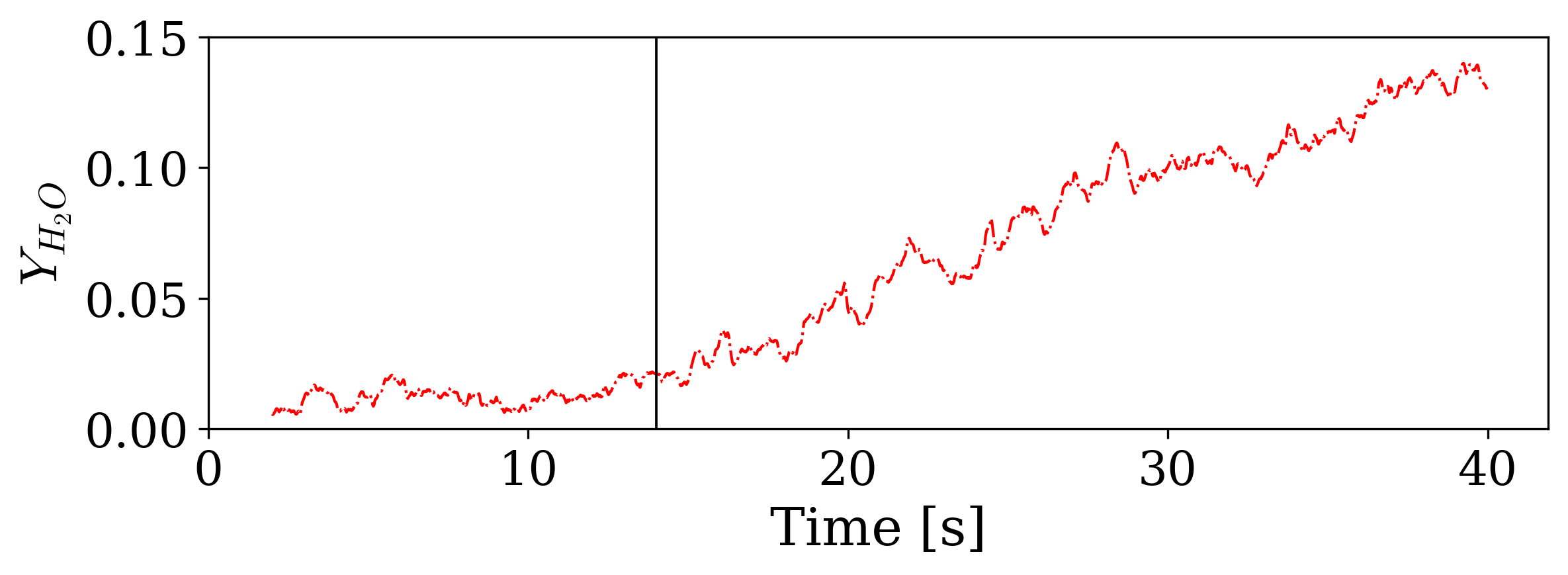}
    \vspace{-1.7em}
    \label{fig:error_log_H2O}
\end{subfigure}

\caption{Time evolution of spatially averaged error $\overline{\Upsilon}_k$ (see~\Cref{sss:error_metric}) for $\textnormal{CO}_2$ and $\textnormal{H}_2\textnormal{O}$ of OpInf with log-transformed mass fractions.}
\label{fig:error_log_transformation}
\end{figure}

In this appendix, we discuss a strategy for learning an OpInf ROM using log-transformed species mass fraction data. As noted in~\Cref{ss:data_pre_processing}, the species mass fractions are not preprocessed in this work, since their values already fall within the target range used in data preprocessing,~[$-1,1$]. Nevertheless, a log-transformation can be applied prior to training to ensure positivity of the predicted mass fractions. Since the minimum mass fraction in the training data is zero (see~\Cref{tab:range_of_variable}), we introduce a small positive tolerance, $0 < \epsilon \ll 1$, to avoid taking the logarithm of zero. The log-transformed mass fractions are then defined as
\begin{equation*}
    Y_{ln} = \ln(Y + \epsilon),    
\end{equation*}
which satisfies $Y_{ln} < 0$ for all training mass fractions in the range $0 \leq Y < 1 - \epsilon$. We then scale the log-transformed mass fractions using the same way as in Eq.~\eqref{eq:scaling}. After training a polynomial OpInf ROM, we propagate the low-dimensional states via the standard OpInf ROM state evolution illustrated in~\Cref{fig:species_limiter_step}. Finally, the high-dimensional states are reconstructed from the predicted low-dimensional states, followed by unscaling and inversion of the logarithm to recover positive mass fractions. The procedure can be summarized as follows:
\begin{enumerate}[noitemsep]
    \item Apply the logarithm to species mass fractions: $Y_{ln} = \ln(Y + \epsilon) < 0$.
    \item Obtain the scaled log-transformed states $Y_{ln}^{\textnormal{scaled}} = \frac{Y_{ln}}{\beta}$, where $\beta = \min \left( Y_{ln} \right) < 0$. This ensures $0 \leq Y_{ln}^{\textnormal{scaled}} \leq 1$.
    \item Train the OpInf ROM.
    \item Propagate the low-dimensional states.
    \item Reconstruct high-dimensional log-transformed mass fractions $\widetilde{Y}_{ln}$.
    \item Obtain the unscaled log-transformed prediction $\widetilde{Y}_{ln}^{'} = \beta \cdot \widetilde{Y}_{ln}$
    \item Recover the mass fraction predictions in the original domain: $Y_{\text{pred}} = e^{\widetilde{Y}_{ln}^{'}} - \epsilon$.
\end{enumerate}
\noindent
Due to the inclusion of $\epsilon$, the recovered mass fraction in step 7 satisfies $Y_{\textnormal{pred}} \geq -\epsilon$. In other words, the predicted mass fractions are bounded below by a negative value infinitesimally close to zero.
In our test, we set $\epsilon = 10^{-16}$.
However, this approach still fails to achieve our goal of physical consistency of mass fractions and stability. While the preprocessed high-dimensional log-transformed mass fractions are strictly bounded ($0 \leq Y_{ln}^{\textnormal{scaled}} \leq 1$ at step~2), the reconstructed log-transformed states $\widetilde{Y}_{ln}$ at step~5 are obtained through the coupled ROM dynamics and are not guaranteed to remain within the same bounds. In particular, if $\widetilde{Y}_{ln} < 0$, then the resultant $\widetilde{Y}_{ln}^{'}$ (after unscaling) may become positive. This, in turn, leads to $Y_\textnormal{pred} > 1$, which violates the physical consistency of mass fractions and may cause instability in long-time ROM predictions.
\Cref{fig:error_log_transformation} shows unstable time-evolution of spatially averaged error $\overline{\Upsilon}_k$ of two variables, $Y_{\textnormal{CO}_2}$ and $Y_{\textnormal{H}_2\textnormal{O}}$.

\bibliography{references}
\bibliographystyle{abbrv}

\end{document}